\documentclass[12pt]{article}

\usepackage{amsmath,amssymb}
\usepackage[dvips]{graphicx,color}
\usepackage[round]{natbib}
\usepackage{mathrsfs}
\usepackage{bm, xcolor}
\usepackage{mathrsfs}
\usepackage{verbatim}
\usepackage{threeparttable}
\usepackage{color}
\usepackage[utf8]{inputenc}
\usepackage{epstopdf}
 \usepackage{booktabs}
 \usepackage{multirow}
\usepackage{graphicx}
\usepackage{subfigure}
\usepackage{float}
\usepackage{listings}
\usepackage{amssymb}
\usepackage{amsthm}
\usepackage{multirow}

\usepackage{caption}
\usepackage{filecontents}

\makeatletter
\def\singlespace{\def\baselinestretch{1}\@normalsize}

\newtheorem{lemma}{Lemma}[section]
\newtheorem{theorem}{Theorem}
\newtheorem{proposition}{Proposition}
\newtheorem{remark}{Remark}
\@addtoreset{equation}{section}
\theoremstyle{definition}
\newtheorem{definition}{Definition}
\renewcommand{\theequation}{\thesection.\arabic{equation}}

\renewcommand{\hat}{\widehat}
\def\hDash{\bot\!\!\!\bot}
\def\singlespace{\def\baselinestretch{1}\@normalsize}

\makeatother

\newcommand{\bS}{\bm{S}}
\newcommand{\bX}{\bm{X}}
\newcommand{\bZ}{\bm{Z}}

\newcommand{\bbeta}{\bm{\beta}}

\newcommand{\btheta}{\bm{\theta}}
\newcommand{\bgamma}{\bm{\gamma}}

\newcommand{\ply}{p_{\lambda_Y}}
\newcommand{\plx}{p_{\lambda_X}}

\newcommand{\mbE}{\mathbb{E}}

\def\newpage{\vfill\eject}

\newdimen\biblioindent    \biblioindent=30pt

 at 7truept

\def\bv{\bm{v}}

\def\beq{\begin{equation}}
\def\eeq{\end{equation}}
\def\beqn{\begin{eqnarray}}
\def\eeqn{\end{eqnarray}}
\def\beqnn{\begin{eqnarray*}}
\def\eeqnn{\end{eqnarray*}}

\def\bM{\bm{M}}
\def\mA{\mathcal{A}}

\begin{document}

\renewcommand{\baselinestretch}{1.3}

\title {\bf Model-Free Statistical Inference on High-Dimensional Data
\footnote{The authors contribute equally and are listed in alphabetical order.}}
\author{Xu Guo$^1$, Runze Li$^2$, Zhe Zhang$^2$, and Changliang Zou$^3$\\
{\small {\small {\it $^1$School of Statistics, Beijing Normal University, China} }}\\
{\small {\small {\it $^2$Department of Statistics, The Pennsylvania State University, USA} }}\\
{\small {\small {\it $^3$School of Statistics and Data Science, Nankai University, China} }}\\
}

\date{February 27, 2022}

\maketitle
\renewcommand{\baselinestretch}{1.5}
\baselineskip=24pt

\noindent {\bf Abstract:}
This paper aims to develop an effective model-free inference procedure for high-dimensional data.
We first reformulate the hypothesis testing problem via sufficient dimension reduction framework.
With the aid of new reformulation, we propose a new test statistic and show that its
asymptotic distribution is $\chi^2$ distribution whose degree of freedom does not
depend on the unknown population distribution. We further conduct power
analysis under local alternative hypotheses. In addition, we study how to control the
false discovery rate of the proposed $\chi^2$ tests, which are correlated, to identify
important predictors under a model-free framework. To this end, we propose a
multiple testing procedure and establish its theoretical guarantees.
Monte Carlo simulation studies are conducted to assess the performance of the proposed tests and
an empirical analysis of a real-world data set is used to illustrate the proposed  methodology.

\par

\vspace{9pt}


\baselineskip=24pt

\pagestyle{plain}
\newpage

\vspace{3mm}

 \section{Introduction}

High-dimensional data are frequently collected in a large variety
of areas such as biomedical imaging, functional magnetic resonance
imaging, tomography, tumor classifications, and finance. Many
regularization methods have been proposed for simultaneous
estimation and variable selection \citep{fan2020statistical}. 
Statistical inference for high-dimensional data receives
considerable attention in the recent literature.
\cite{zhang2014confidence} proposed the debiased Lasso, a
low-dimensional projection  approach to construct confidence
interval and perform hypothesis testing, for linear regression
models. \cite{van2014asymptotically} and
\cite{javanmard2014confidence} proposed the desparsified Lasso
under the setting of generalized linear models, and
\cite{ning2017general} proposed a decorrelated score method that
can be applied to a more general family of penalized M-estimators.
The debiased Lasso, desparsified Lasso and the decorrelated score
method share the same spirit. See, for example, Chapter 7 of
\cite{fan2020statistical} for detailed discussion.
\cite{fang2017testing} generalized the decorrelation method for
the low dimensional hypothesis testing approach in
high-dimensional proportional hazards models. \cite{fang2020test}
extended the decorrelated score method to longitudinal data with
ultrahigh-dimensional predictors. \cite{shi2019linear} developed
the constrained partial regularization method for testing linear
hypotheses in high-dimensional generalized linear models.
\cite{sun2021targeted} proposed a modified profile
likelihood-based statistic for hypothesis testing without
penalizing the parameters of interest. Aforementioned statistical
inference procedures were developed for a specific model such as linear model, generalized linear model or Cox's
model. This work aims to develop a model-free statistical
inference procedure for high-dimensional data.

In the initial stage of high-dimensional data modeling, it may be
quite challenging in correctly specifying a regression model, if
not impossible. Furthermore there is lack of effective procedures
to validate specified model assumption in the high-dimensional
setting. Thus, it is of great interest to develop statistical
inference procedures for high-dimensional data without a
pre-specified parametric model.
%
%
To be precise, let $Y\in\mathbb{R}$ be the response variable along with predictor vector
$\bX=(X_1,\cdots,X_p)^{\top}\in\mathbb{R}^p$. Denote $F(Y |\bX)$ to be the conditional distribution of $Y$ given $\bX$. The active
set $\mA$ is defined as
$$
\mA=\{j: F(Y|\bX)\,\,\mbox{functionally depends on $X_{j}$}\}.
$$
Then $\mA^c$, the complement of $\mA$, is the index set consisting
of all irrelevant predictors. Let $\bX_{\mA}=\{X_{j}, j\in\mA\}$
denote the vector containing all the active predictors. Thus, $Y$
is independent of $\bX$ given $\bX_{\mA}$, denote it by $Y\hDash
\bX|\bX_{\mA}$. It is of interest to test whether a predictor
$X_{j}$ contributes to the response $Y$ or not given the other
predictors. This can be formulated as testing the hypothesis
whether $H_{0j}: j\in\mA^c.$

In this paper, we propose a new model-free test for $H_{0j}$ in
the presence of ultrahigh dimensional predictors. We first develop
a new reformulation of $H_{0j}$ by using statistical framework of
sufficient dimension reduction \citep{li2018sufficient}.
Based on this reformulation, we further propose a test statistic
for $H_{0j}$ and show that it converges to $\chi^2$ distribution
whose degree of freedom does not depend on unknown population
quantities. We also show that the newly proposed test statistic
can retain power under local alternative hypotheses.

Under the model-free framework, we study the false discovery rate (FDR) control in
the large scale simultaneous testing problem: $H_{0j}: j\in\mA^c\,\,\mbox{for all}\,\,j=1,\cdots,p.$
Classical FDR control procedures
including Benjamini-Honchberg (BH) procedure \citep{benjamini1995controlling} and
Storey procedure \citep{storey2002direct}
are typically
built on the independence assumption.  Other FDR control procedures such as Benjamini-Yekutieli (BY)
procedure \citep{benjamini2001control} require positive dependence structure and tend to be
overly conservative \citep{cai2019large}. These procedures may fail to control the FDR
in the presence of complex and strong dependence structure in practice.
\cite{liu2013gaussian} developed FDR control procedures for multiple $z$-tests under Gaussian
graphical model setting by assuming the number of pairs of strongly correlated tests bounded.
This method can be viewed as a truncated version of BH method, but it cannot be directly extended
to the underlying setting of this paper because of complicated dependence structure.
We propose a new FDR control procedure by using the maximum correlation coefficient
to measure the dependence of two tests, and show that the proposed procedure asymptotically controls the FDR at
nominal level under mild conditions.

The paper is organized as follows. In section 2, we present a new reformulation. In section 3,
we propose a new test procedure and investigate its asymptotic properties
under the null and local alternative hypotheses. In section 4, we investigate the FDR control procedure which is important to identify active predictors.
We present numerical studies in section 5 and conclusions in section 6. Proofs are given in the Appendix.
Technical lemmas and their proofs are given in the supplementary material of this paper.

\section{A reformulation via sufficient dimension reduction}

We now reformulate the hypothesis $H_{0j}: j\in\mA^c$ by using concepts related to
sufficient dimension reduction.
Note that $H_{0j}$ is called marginal coordinate hypothesis in
the seminal work of \cite{cook2004testing}, in which the author evaluated the
predictor's effect in a model-free setting by adopting the
sufficient dimension reduction framework \citep{li2018sufficient}.
Following \cite{cook2004testing}, several authors have developed test procedures for
$H_{0j}$ \citep{shao2007marginal, yu2016model, dong2016permutation} when the dimension $p$ is
fixed. These tests have a major limitation that the asymptotic distributions of these test statistics are
sums of weighted $\chi^2$ distributions with unknown weights. This makes their practical implementation challenging.

Define the central subspace
$\mathcal{S}_{Y|\bX}$ to be the minimum subspace $\mathcal{S}$ given
by the column space of a $p\times d$ matrix
$\boldsymbol{B}=(\bbeta_1,\cdots,\bbeta_d)$ with $d<p$ such that $Y\hDash
\bX|\boldsymbol{B}^{\top}\bX.$ Under some mild conditions, the central subspace
uniquely exists and contains all information of $F(Y |\bX)$
\citep{
li2018sufficient}. Throughout this paper, it is assumed that
$\mathcal{S}_{Y|\bX}$ exists and its dimension $d$ is fixed.

Based on the concepts of the central subspace, we can reformulate
$H_{0j}$. To this end, we firstly consider the estimation of the
central subspace $\mathcal{S}_{Y|\bX}$ because the basis matrix
$\boldsymbol{B}$ is unknown.  Without loss of generality, assume that
$\mbE(\bX)=\boldsymbol{0}$, and $\boldsymbol{\Sigma}=\mbox{Var}(\bX)>0$. Following the literature
of sufficient dimension reduction, we impose linearity condition (LC):
$$
\mbE[\bX|\boldsymbol{B}^{\top}\bX]\,\,\mbox{is a linear function of}\,\,\boldsymbol{B}^{\top}\bX.
$$
The LC is satisfied when $\bX$ follows an elliptical distribution,
but can be more general in asymptotical sense
\citep{hall1993almost}. Note that this condition is imposed only on the distribution of $\bX$,
rather than the conditional distribution $F(Y|\bX)$. In fact, the relationship between $Y$ and $\bX$ is
 unspecified, and hence is model-free.


Under the linearity condition, it follows by \cite{yin2002dimension} that for any function $f(Y)$,
$$
\boldsymbol{\Sigma}^{-1}\mbox{Cov}(\bX,f(Y))\in \mathcal{S}_{Y|\bX}.
$$
This enables us to choose a series of transformations
on $Y$: $f_1(Y), \cdots, f_h(Y)$, with pre-specified $h>d$. We will discuss how to choose $f_k(\cdot)$ and $h$ in
Section 5.
Define
$$
\bbeta_k^0=\arg\min_{\bbeta_k}\mbE[\{f_k(Y)-\boldsymbol{X}^{\top}\bbeta_k\}^2],\quad \mbox{for}\quad k=1,\cdots,h.
$$
Denote $\boldsymbol{B}_0=(\bbeta_1^0,\cdots, \bbeta^0_h)$.
Then $\mbox{Span}(\boldsymbol{B}_0)\subseteq\mathcal{S}_{Y|\bX}$. Following
the literature of sufficient dimension reduction, we take
one step further by assuming the coverage condition (CC)
$$
\mbox{Span}(\boldsymbol{B}_0)=\mathcal{S}_{Y|\bX},
$$ whenever
$\mbox{Span}(\boldsymbol{B}_0)\subseteq\mathcal{S}_{Y|\bX}$ so that the subspace spanned by $\boldsymbol{B}_0$ coincides with the central subspace. This condition is reasonable. Actually the larger $h$ is, the $\mbox{Span}(\boldsymbol{B}_0)$ is larger. Thus when $h$ is large, $\mbox{Span}(\boldsymbol{B}_0)$ could be very close to $\mathcal{S}_{Y|\bX}$.

For $j=1,\cdots,p$, let $\boldsymbol{b}_j^{\top}= (\beta^0_{1j},\cdots,\beta^0_{hj})$
be the $j$th row of $\boldsymbol{B}_0$, where $\beta^0_{kj}$ is the $j$th element of
$\bbeta^0_k\in\mathbb{R}^p,k=1,\cdots,h$.
Note that if $j\in\mA^c$, then any of the $h$ linear combinations $\bX^{\top}\bbeta_k^0, k=1,\cdots,h$
must not involve $X_{j}$. Thus, $Y\hDash \bX|\bX_{\mA}$ and $Y\hDash \bX|\boldsymbol{B}_0^{\top}\bX$
imply that $\sum_{k=1}^h |\beta^0_{kj}|>0$ for $j\in\mA$, and $\sum_{k=1}^h |\beta^0_{kj}|=0$ for $j\in\mA^c$.
Thus, the following proposition are valid.
\begin{proposition}
Suppose that the LC and CC conditions hold, we have:
$H_{0j}: j\in\mA^c$ if and only if $H_{0j}': \boldsymbol{b}_j= \boldsymbol 0$.
\end{proposition}
This reformulation converts the original hypothesis testing problem into a parametric hypothesis without
knowing the specific form of the conditional distribution of $Y$ given $\bX$.
We next construct a score-type statistic for testing $H_{0j}'$.

\section{A new test statistic and its limiting distribution}

In this section, we develop a test statistic for   $H_{0j}'$ to achieve the goal of testing
$H_{0j}$. To get the insight of proposed test for $H_{0j}'$, we first consider the low dimensional setting in which $p$ is fixed. By the definition of $\bbeta_k^0$, we may consider using $\mbE
[X_{j}\{f_k(Y)-\bX^{\top}\bbeta_k^0\}]$ (i.e., the derivative of $\mbE[\{f_k(Y)-\bX^{\top}\bbeta_k^0\}^2]$ with respect to $\beta_{kj}^0$), for $k=1,\cdots, h$ to construct a test statistic. This is similar to the score test in likelihood setting. Under $H_{0j}'$,
\[
\mbE[X_{j}\{f_k(Y)-\bX^{\top}\bbeta_k^0\}]=\mbE[X_j\{f_k(Y)-\bZ_j^{\top}\bgamma_{kj}\}],
\]
where $\bZ_j$ is the subvector of $\bX$ without $X_{j}$, and $\bgamma_{kj}$ is the subvector
of $\bbeta_{k}^0$ without $\beta_{kj}^0$.

Suppose that $\{\bX_i, Y_i\}$, $i=1,\cdots, n$, are random samples generated from $\{\bX, Y\}$. Similarly, we denote $\bZ_{ij}$ for the sample  of $\bZ_j$. By the definitions of $\bbeta_k^0$, we then obtain the least squares estimate of $\bgamma_{kj}$, $\hat\bgamma_{kj}$, by a linear regression of $f_k(Y)$ on $\bZ_j$.
Then it is natural to consider the score-type test statistic:
$$T_{nk}^j=\frac{1}{\sqrt n}\sum_{i=1}^n X_{ij}(f_k(Y_i)-\bZ^{\top}_{ij}\hat\bgamma_{kj}).$$
It can be shown that under the null hypothesis, $(T_{n1}^j,\cdots,T_{nh}^j)$ jointly converges to a multivariate normal distribution with
zero mean.

We now consider high-dimensional setting.
The test procedure does not work in high-dimensional setting since the least squares estimate is not well defined.
However, we may consider the score type test $T_{nk}^j$ by replacing the least squares estimate
with the corresponding penalized estimate $\hat\bgamma_{kj}$ of $\bgamma_{kj}$.
In high-dimensional setting, it is not  uncommon to impose sparsity assumption
on regression coefficients. That is, only  a small
subset of the predictors are significant to the response variable.
We estimate $\boldsymbol{\bgamma}_{kj}$ by its penalized least squares estimate
\begin{equation}
\hat{\bgamma}_{kj}=\arg\min\frac{1}{2n}\sum_{i=1}^n \{f_k(Y_i)-\bZ^{\top}_{ij}\bgamma_{kj}\}^2
+\sum_{l=1}^{p-1}\ply(|\gamma_{kj,l}|),\,\,    k=1,\cdots,h,
    \label{gamma1}
\end{equation}
where $\ply(\cdot)$ is a penalty function with a tuning parameter $\lambda_Y$.

\begin{remark} For multiple testing problem to be studied in next section, we need to calculating all
$\hat{\bgamma}_{kj}$, $j=1,\cdots, p$ and $k=1,\cdots, h$. This requires to
solve $hp$ regularized optimization problems, and leads to expensive computational cost.
For multiple testing problem, we recommend to estimate $\bbeta^0_k$ by using
\begin{equation}
\hat{\bbeta}_{k}=\arg\min\frac{1}{2n}\sum_{i=1}^n \{f_k(Y_i)-\bX^{\top}_{i}\bbeta_{k}\}^2+\sum_{l=1}^{p}\ply(|\beta_{kl}|),\,\,
k=1,\cdots,h.
\label{gamma2}
\end{equation}
Then set $\hat{\bgamma}_{kj}$ as the subvector of $\hat{\bbeta}_{k}$ without $\hat\beta_{kj}$. This can
save computational cost dramatically.
\end{remark}

The test statistic $T_{nk}^j$ can be further refined since it could fail in some situation. To
see this, note that
\begin{eqnarray*}
T_{nk}^j&=&\frac{1}{\sqrt n}\sum_{i=1}^n X_{ij}(f_k(Y_i)-\bZ^{\top}_{ij}\hat\bgamma_{kj})\\
&=&\frac{1}{\sqrt n}\sum_{i=1}^nX_{ij}(f_k(Y_i)-\bZ^{\top}_{ij}\bgamma_{kj})+\frac{(\bgamma_{kj}-\hat\bgamma_{kj})^{\top}}{\sqrt n}\sum_{i=1}^n \bZ_{ij}X_{ij}.
\end{eqnarray*}
Although under mild conditions, the first term converges to a normal distribution,
the second term may not be asymptotically normal. Indeed, for the penalized estimate
of $\bgamma_{kj}$, it may not be always valid without strong condition on $\bgamma_{kj}$, such as minimal signal
condition \citep{fan2020statistical}, to ensure that
$\sqrt n(\hat\bgamma_{kj}-\bgamma_{kj})$ converges to normal distribution.
Moveover, the Euclidean norm of the term $n^{-1}\sum_{i=1}^n \bZ_{ij}X_{ij}$ can diverge since its dimension
is $p$. Thus it is of importance to refine the test statistic $T_{nk}^j$ so that it performs well in general
settings. We next employ the idea of orthogonalization to refine $T_{nk}^j$.
The orthogonalization plays a critical role in reducing the bias from the estimators of
high-dimensional nuisance parameters \citep{Belloni.2015, ning2017general, Belloni.2018}.

Instead of employing $\mbE[X_j\{f_k(Y)-\bZ_j^{\top}\bgamma_{kj}\}]$ in high-dimensional setting, we propose
using
\beq
\mbE\Big[ (X_{j}-\bZ_j^{\top}\btheta_j^*)\{f_k(Y)-\bZ_j^{\top}\bgamma_{kj}\}\Big]
\label{o1}
\eeq
to construct a test statistic for $H_{0j}'$, where $\btheta_j^*=\mbE[\bZ_j\bZ_j^{\top}]^{-1}\mbE[\bZ_jX_{j}]$.
This is because the term in (\ref{o1}) possesses the orthogonality property:
$$
\frac{\partial}{\partial\bgamma_{kj}}\mbE\Big[ (X_{j}-\bZ_j^{\top}\btheta_j^*)\{f_k(Y)-\bZ_j^{\top}\bgamma_{kj}\}\Big]
=\mbE\left[\bZ_{j}\left(X_{j}-\bZ^\top_{j}\btheta_j^*\right)\right]=0
$$
under some conditions.

In order to reduce the influence of nuisance parameters, we estimate $\btheta_j^*$ by a penalized least squares
estimate
\begin{equation}
\hat{\btheta}_j=\arg\min\frac{1}{2n}\sum_{i=1}^n (X_{ij}-\bZ^{\top}_{ij}\btheta_j)^2+\sum_{l=1}^{p-1}\plx(|\theta_{j,l}|),
\label{theta}
\end{equation}
where $\plx(\cdot)$ is a penalty function with a tuning parameter $\lambda_X$.

Define  $\boldsymbol{S}_i^j=(S_{i1}^j,\cdots,S_{ih}^j)^{\top}$ with
$
S_{ik}^j= (X_{ij}-\bZ^{\top}_{ij}\hat{\boldsymbol{\theta}}_j)(f_k(Y_i)-\bZ^{\top}_{ij}\hat{\bgamma}_{kj})
$
for each $k$.
By the orthogonality property of $\mbE\Big[ (X_{j}-\bZ_j^{\top}\btheta_j^*)\{f_k(Y)-\bZ_j^{\top}\bgamma_{kj}\}\Big]$,
it is natural to construct a test statistic based on
$$
\bar{\boldsymbol{S}}_n^j = \frac{1}{\sqrt{n}}\sum_{i=1}^n \boldsymbol{S}_i^j.
$$
That is, $\bar{\boldsymbol{S}}_n^j = (\bar{S}_{n1}^j,\cdots,\bar{S}_{nh}^j)^{\top}$ with
$
\bar{S}_{nk}^j=n^{-1/2}\sum_{i=1}^n (X_{ij}-\bZ^{\top}_{ij}\hat{\boldsymbol{\theta}}_j)(f_k(Y_i)-\bZ^{\top}_{ij}\hat{\bgamma}_{kj}).
$



As shown in the Appendix, $\bar{\boldsymbol{S}}_n^j$ asymptotically follows
a normal distribution with mean 0 and covariance matrix $\boldsymbol{\Omega}_j$, where the $(l,k)$-element of
$\boldsymbol{\Omega}_j$ is $\mbE[\eta_l^j\eta_k^j(X_{j}-\bZ^{\top}_{j}\boldsymbol{\theta}^*_j)^2]$
for $l,k=1,\cdots,h$, where $\eta_k^j=f_k(Y)-\bZ_j^{\top}\bgamma_{kj},k=1,\cdots,h$.
Thus, we propose the following test statistic for $H_{0j}'$
\begin{align*}
W_{nj}=\bar{\boldsymbol{S}}^{j\top}_n\hat{\boldsymbol{\Omega}}_j^{-1}\bar{\boldsymbol{S}}_n^j,
\end{align*}
where $\hat{\boldsymbol{\Omega}}_j$ is a consistent estimate of $\boldsymbol{\Omega}_j$ with
$(k,l)$-elements $\hat{\omega}_{kl}$ being
\[
\frac{1}{n}\sum_{i=1}^n(f_l(Y_i)-\bZ^{\top}_{ij}\hat{\bgamma}_{lj})(f_k(Y_i)-\bZ^{\top}_{ij}\hat{\bgamma}_{kj})
(X_{ij}-\bZ^{\top}_{ij}\hat{\boldsymbol{\theta}}_j)^2.
\]

Let $s_Y$ be the sparsity level for $\bgamma_{kj},k=1,\cdots,h$.
That is, $s_Y=\max\limits_{1\le j\le p,1\le k\le h} \|\bgamma_{kj}\|_0$.
Also denote the sparsity level for the parameter $\btheta^*_j$ to
be $s_X=\max\limits_{1\le j\le p}\|\btheta_j^*\|_0$. Before we introduce
technical conditions, let us introduce some new notations related
to sub-Gaussian and sub-exponential random variables.

\begin{definition}
(i) A random variable $X$ is sub-Gaussian if its tail probability
satisfies $\Pr\{|X| \geqslant t\} \leqslant 2 \exp \left(-t^{2} /
K_{1}^{2}\right)$ for all $t \geqslant 0$.
$\|X\|_{\varphi_2}=\inf\{t>0: \mbE\exp(X^2/t^2)\leqslant 2\}$ is
the sub-Gaussian norm of $X$, and is the smallest $K_1$.

(ii) A random variable $X$ is sub-exponential if its tail
probability satisfies $\Pr\{|X| \geqslant t\} \leqslant 2 \exp
\left(-t / K_2\right)$ for all $t \geqslant 0$.
$\|X\|_{\varphi_1}=\inf\{t>0: \mbE\exp(|X|/t)\leqslant 2\}$ is the
sub-exponential norm of $X$, and is the smallest $K_2$.
\end{definition}

The following technical conditions are imposed to establish the asymptotical null distribution of $W_{nj}$,
although they may not be the weakest conditions.
\begin{itemize}
\item[(B1)] $\|\hat{\bgamma}_{kj}-\bgamma_{kj}\|_1=O_p(\lambda_Ys_Y), \|\hat{\boldsymbol{\theta}}_j-\boldsymbol{\theta}^*_j\|_1=O_p(\lambda_Xs_X)$.
\item[(B2)] $\eta_k^j,k=1,\cdots,h,$ and $ X_{j},j=1,\cdots,p$ are all sub-Gaussian random variables
with finite expectation. $\left\|\eta_{k}^j\right\|_{\psi_{2}}$ and $\left\|X_{j}\right\|_{\psi_{2}}$ are uniformly bounded.
\item[(B3)] $\max\{s_Y, s_X\}=o(\sqrt n/\log p)$, $\max\{\lambda_X,\lambda_Y\} = C\sqrt{\log p/ n}$.
\item[(B4)] $\boldsymbol{\Sigma}=\mbox{Cov}(\boldsymbol{X})>0$, and $\boldsymbol{\Omega}_j>0$. 

\end{itemize}

These conditions are commonly used in the literature. Condition
(B1) is for error bound of the penalized least squares estimators
in linear regression models. The penalized least squares estimates
with the Lasso, SCAD and MCP penalties satisfy this condition
\citep{fan2020statistical}. Condition (B2) is imposed for
establishing the $L_{\infty}$ norm of
$n^{-1/2}\sum_{i=1}^n\eta_{ik}^j\bZ_{ij}$. Sub-Gaussian assumption
is a commonly used condition in the literature
\citep{fan2020statistical}. 
Furthermore, the dimension of $p$ is allowed to be an exponential
order of the sample size $n$ according to Conditions (B2) and
(B3). The rate for the sparsity levels in Condition (B3) is
commonly imposed, see for instance \cite{ning2017general}.
Condition (B4) is also a mild condition.

The limiting distribution of $W_{nj}$ under the null hypothesis
and local alternative hypotheses are given in the following
theorem, whose proof is given in the Appendix.

\begin{theorem}\label{thm1}
Suppose that the LC and CC conditions, and Conditions (B1)|(B4) are valid.

\begin{description}
\item{(i)} Under the null hypothesis $H_{0j}$, it follows that
$
W_{nj}\rightarrow \chi^2_h
$
in distribution, where $\chi^2_h$ stands for
the chi-square distribution with $h$ degrees of freedom;

\item{(ii)} Under the alternative hypothesis $H_{1n}: \sqrt{n}
\boldsymbol{b}_{j} \rightarrow \boldsymbol{b}\neq \boldsymbol{0}$,
it follows that
$$
W_{nj} \rightarrow \chi_{h}^{2}\left(\delta_j^{2} \boldsymbol{b}^{\top} \boldsymbol{\Omega}^{-1}_j \boldsymbol{b}\right)
$$
in distribution,
where $\chi_{h}^{2}(a)$ is the non-central chi-square distribution with $h$ degrees of freedom and
noncentrality parameter $a$, and
$\delta_j=\mbE(X_{j}^{2})-\mbE(X_{j} \bZ_{j}^{\top})
\mbE(\bZ_{j} \bZ_{j}^{\top})^{-1} \mbE(\bZ_{j} X_{j})$.
\end{description}
\end{theorem}
Theorem~\ref{thm1}(a) implies that
we are able to test the hypothesis whether $F(Y|\bX)$ functionally depends on the $j$th predictor or not
in ultrahigh dimensional setting without imposing a specific functional form on regression model.
This desirable feature is significantly different from the existing works based on
specific parametric model such as linear model or generalized linear model.

Theorem~\ref{thm1}(b) implies that the proposed test  $W_{nj}$ can
detect the alternative hypotheses which converge to the null at
the rate of $n^{-1 / 2}$. From the proof of Theorem~\ref{thm1}(b),
we can show that $W_{nj} \rightarrow \infty$ for fixed alternative
hypothesis $H_{1}: \boldsymbol{b}_{j} \neq \boldsymbol{0}$. Thus,
$W_{n}$ has asymptotic power 1 for any fixed alternative
hypothesis.

\section{False discovery rate control}
In Section 3, we propose a test of significance for an individual
predictor. In practice, it is of interest to test simultaneously
all $p$ hypotheses. That is, $H_{0 j}: j \in \mathcal{A}^{c}$ for
all $j=1, \cdots, p$.  This is potentially useful to identify
important features in the high-dimensional data. To avoid spurious
discoveries, the FDR control is commonly adopted. Recently,
\cite{bc2015} introduced a novel knockoff framework to control the
FDR under the linear model. \cite{candes2018panning} further
developed a model-X knockoff procedure that can control FDR
without assuming a specific regression model. However, the model-X
knockoff procedure requires knowing the joint distribution of the
predictors. This makes the knockoff method challenging in
practical implementation. See also \cite{fan2019rank}. This
section aims to develop a FDR control procedure under a model-free
framework. The new procedure does not require the knowledge of the
joint distribution of the predictors.

For a series of individual hypothesis $H_{0j}$, we can construct the corresponding test statistics $W_{nj}$. As shown in the Theorem 1, under null hypothesis $W_{nj}$ follows an asymptotically chi-squared distribution with degrees of freedom $h$.

A FDR controlling procedure is to find a threshold $t$ to control the multiple testing effect. For any threshold  $t>0$, let $R_{0}(t)=\sum_{j \in \mathcal{A}^{c}} I\left(W_{nj}\geqslant t\right)$, and $R(t)=\sum_{j=1}^p I\left(W_{nj} \geqslant t\right)$ be the total number of false discoveries and the total number of discoveries associated with $t$,  respectively.

Define false discovery proportion (FDP) and FDR as follows:
$$
\mathrm{FDP}(t)=\frac{R_{0}(t)}{R(t) \vee 1}, \quad \mathrm{FDR}(t)=\mbE\{\mathrm{FDP}(t)\}.
$$
Under a pre-specified FDR level $\alpha$, the ideal choice of the threshold $t$  would be
$$
t_{0}=\inf \left\{t\in \mathbb{R}: \operatorname{FDP}(t) \leqslant \alpha\right\}.
$$
In practice,  we need to estimate $R_0(t)$ since the inactive set is unknown. Intuitively,
$R_0(t)$ can be reasonably estimated by $|\mathcal{A}^c|G(t)$, where $G(t)=\Pr(\chi^2_h\geqslant t)$. As a result, $\mbox{FDP}(t)$ can be estimated by
$$
\widehat{\mbox{FDP}}(t) = \frac{|\mathcal{A}^c|G(t)}{R(t) \vee 1}.
$$
Thus we propose to estimate the threshold level $t$ by
$$
\hat{t} = \inf \left\{0 \leqslant t \leqslant b_p: \widehat{\operatorname{FDP}}(t) \leqslant \alpha\right\}.
$$
and reject the null hypothesis $H_{0j}$ if $W_{nj} \geqslant \hat{t}$, where $b_p=2 \log p + 2d_0\log\log p$
with $2d_0<h-4$. If there is no $t\in [0, b_p]$ such that
$\widehat{\operatorname{FDP}}(t) \leqslant \alpha$, we will directly set $\hat{t} = 2\log p+ (h-1)\log \log p$.

In above, we estimate $R_0(t)$ by $|\mathcal{A}^c|G(t)$ when $t
\in [0, b_p]$. In fact under the restriction $2d_0 < h-4$ and
other mild conditions, we will show
\begin{equation}
\sup _{0 \leqslant t \leqslant b_p}\left|\frac{\sum_{j \in \mathcal{A}^{c}} I\left(W_{nj} \geqslant t\right)}{|\mathcal{A}^c|G(t)}-1\right| \rightarrow 0,
\label{e4.1}
\end{equation}
in probability. In other words, it implies that the total number
of false discoveries $R_0(t)$ can be consistently estimated by
$|\mathcal{A}^c|G(t)$ in this pre-specified region. In practice,
$|\mathcal{A}^c|$ is not a prior knowledge, we could approximate
$|\mathcal{A}^c|$ by $p$ since $|\mathcal{A}| = o(p)$.


\begin{remark}
The choice of $d_0$ depends on the value of $h$. Since $h-4$ is
the boundary value to guarantee the theoretical results, we
suggest choosing a slightly smaller value $d_0 = h-4-\epsilon_h$
in practice, where $\epsilon_h$ is a small positive constant. It
is noteworthy that $d_0$ is allowed to be negative if $h\leqslant
4$. In \cite{xia2018joint}, they studied the case of asymptotic
normal test statistic which corresponds to $h=1$. They set the
search region to be $[0,\sqrt{2\log p - 2\log \log p}]$, where the
right boundary point is smaller than $\sqrt{2\log p}$.
\end{remark}


It is crucial to characterize the dependence structure across the tests in order to establish (\ref{e4.1}).
Denote
$$
\widetilde{\bS}_{n}^j=\frac{1}{\sqrt{n}} \sum_{i=1}^{n} \boldsymbol{\eta}_{i}^j\left(X_{i j}-\bZ_{i j}^{\top} \boldsymbol{\theta}^*_j\right), \quad
\widetilde{W}_{n j} = \widetilde{\boldsymbol{S}}_{n}^{j \top} \boldsymbol{\Omega}_j^{-1} \widetilde{\boldsymbol{S}}_n^j.
 $$
Further denote $\widetilde{\boldsymbol{K}}_n^j = \boldsymbol{\Omega}_j^{-\frac{1}{2}}\widetilde{\boldsymbol{S}}_n^j$. Define the maximum correlation coefficient
$$
\rho_{i j}^{*}=\max _{\|\boldsymbol{a}\|=1,\|\boldsymbol{b}\|=1}\left|\operatorname{corr}\left(\boldsymbol{a}^{\top}
\widetilde{\boldsymbol{K}}_n^i, \boldsymbol{b}^{\top}\widetilde{\boldsymbol{K}}_n^j \right)\right|
$$
which quantifies the dependence between $\widetilde{W}_{n i}$ and $\widetilde{W}_{n j}$. For $i,j\in \mathcal{A}^c$,
 define
$$
\mathcal{B}_1:=\left\{(i,j): i \neq j, \rho_{i j}^{*}\geqslant(\log p)^{-2 - \epsilon_0}\right\} \text{for some $\epsilon_0 > 0$}
$$
which contains the pairs $(i,j$) with high correlation.

In order to ensure the estimated FDR could be controlled at pre-specified level,
the difference between constructed tests $W_{nj}$ and $\widetilde{W}_{nj}$ must be uniformly bounded.
The following proposition implies that this is valid under mild assumptions

\begin{proposition}
Suppose that the LC and CC conditions, and conditions B1, B2 and B4 are valid.
Assume $\lambda_X, \lambda_Y  =O(\sqrt{\frac{\log p}{n}})$, $\max(s_X, s_Y) = s$. It follows that
\begin{itemize}
\item[(i)] $\|\bar{\boldsymbol{S}}_n^j-\widetilde{\boldsymbol{S}}_n^j\|_1 = O_p(F(n,p,s)) = O_p(\frac{\log p}{\sqrt{n}}s)$,
\item[(ii)] $\|\hat{\boldsymbol{\Omega}}_j -  \boldsymbol{\Omega}_j\|_2 =  O_p(G(n,p,s))$,
\end{itemize}
hold uniformly for $j \in \mathcal{A}^c$, where
$$
G(n,p,s)  = \sqrt{\frac{(\log np)^{5}}{n}}\vee \frac{s\log p\log (np)}{n}\vee
s^2\log (np) (\frac{\log p}{n})^{\frac{3}{2}}
\vee s^3\log (np) (\frac{\log p}{n})^2.
$$
Furthermore, it follows that
$$
\max\limits_{j\in \mathcal{A}^c}|W_{nj} -\widetilde{W}_{nj}| = O_p\left(\max\left(\frac{(s\log p)^2}{n}, \frac{s(\log p)^2}{\sqrt{n}},G(n,p,s)\log p\right)\right).
$$
\label{prop1}
\end{proposition}
The proof of this proposition is given in the supplementary material of this paper.
We next show the FDR control result of the proposed procedure in Theorem~\ref{thm2} below.
The following technical conditions are imposed to facilitate its technical proof.

\begin{itemize}
\item[(D1)] There exists a positive constant $\epsilon_1$ such that $|\mathcal{B}_1| = O([\log p]^{h-2d_0-2-\epsilon_1})$.
\item[(D2)]  $\max \left(F(n,p,s),G(n,p,s)\log p \right) = o(1)$.
\item[(D3)]$\lambda_{\min}(\boldsymbol{\Omega}_j^{-1}) \geqslant \lambda_0>0$ and $\lambda_{\max}(\boldsymbol{\Omega}_j^{-1})\leqslant \lambda_1$ uniformly in $j = 1,\cdots,p$.
\item[(D4)] $\delta_j\geqslant c>0$ uniformly for $j$ where $\delta_j=\mbE(X_{j}^{2})-\mbE(X_{j} \bZ_{j}^{\top})
\mbE(\bZ_{j} \bZ_{j}^{\top})^{-1} \mbE(\bZ_{j} X_{j})$.
\end{itemize}

These conditions are mild and commonly used in the literature of
FDR. See, e.g.,  \cite{liu2013gaussian},
\cite{xia2018joint} and references therein. 
Condition D1 requires a bounded number of strongly correlated pairs which correspond to tests in inactive set.
The purpose is to control the variance of $R_0(t)$. Condition D2 assumes the difference between
constructed test and its ideal version is uniformly bounded with rate $o(1)$, and therefore
to ensure the consistency of $\sum_{j \in \mathcal{A}^{c}} I\left(W_{nj} \geqslant t\right)$
to $|\mathcal{A}^c|G(t)$. 
Condition D3 assumes the smallest and the largest eigenvalues of $\boldsymbol{\Omega}_j$ are uniformly bounded.

\begin{theorem}
 Under Conditions in Proposition 2 and Conditions D1--D4 with $p \leqslant c n^{r}$ for some $c>0$ and $r>0$. In addition, assume $|\mathcal{A}|=o(p)$. It follows
that
$$
\limsup _{(n, p) \rightarrow \infty} \operatorname{FDR}(\hat{t}) \leqslant \alpha
$$
with probability $1$.
\label{thm2}
\end{theorem}
In Theorem 1, the dimension $p$ is allowed to grow exponentially
fast in $n$. However, in order to control FDR,  it requires $p
=O(n^r)$ to apply the moderate deviation result in
\cite{liu2013gaussian}.

\section{Numerical studies}

In this section, we conduct numerical studies to assess the finite
sample performance of the proposed procedures. To implement the
proposed test, we need to determine the transformation functions
$f_k(\cdot)$. In our simulation, we take the $f_k(\cdot),
k=1,\cdots, h$ to be a set of linear B-spline bases with $h-2$
inner knots, which yields a set of $h$ linear B-spline bases. In
all simulation studies, we set $p=2000$, and all simulation results
are based on 1000 replications. We consider the Lasso
\citep{tibshirani1996regression} and the SCAD
\citep{fan2001variable} in the penalized least squares estimate
introduced in Section 3. We first study the impact of $h$ on the
performance of the proposed test.

\subsection{The choice of $h$}

One needs to choose the value of $h$ to implement the proposed procedure.
We investigate this issue by Monte Carlo
study. To this end, we generate the data from the following three models.

Model I: $Y=X_1+X_2+\epsilon.$

Model II:
$Y=(X_1+X_2)/\{0.5+\left(1.5+X_3+X_4\right)^{2}\}+0.1\epsilon.$

Model III: $Y=3\sin(X_1)+3\sin(X_p)+\exp(-2X_3)\epsilon.$

These three models correspond to linear model, nonlinear model and
heteroscedastic model, respectively. The dimension of central
subspace, $d$, is 1, 2, and 3, respectively. The covariate vector
$\bX=(X_1,\cdots, X_p)$ is generated from $N({\bf 0},\Sigma)$ with
the $(i,j)$-element of $\Sigma$ being $\sigma_{ij}=0.5^{|i-j|}$
for $1\leq i,j\leq p$. The error $\epsilon$ is generated from
$N(0,1)$
and independent of $\bX$. The active sets $\mA$ for these models are $\{1, 2\}, \{1, 2,3, 4\}$, and $\{1, 3, p\}$,
respectively. We consider sample size $n=200$ and $400$.

To examine the impact of $h$ on the performance of $W_{nj}$, we
set $h=1,2,\cdots,20$ and plot the power curve as a function of
$h$ for several $j$s. This enables us to see the overall trend of
the power function as $h$ increases. Moreover, we study the impact
of choice of penalty function on the performance of test
statistics.

\begin{figure}
    \centering
    \includegraphics[width=1\textwidth]{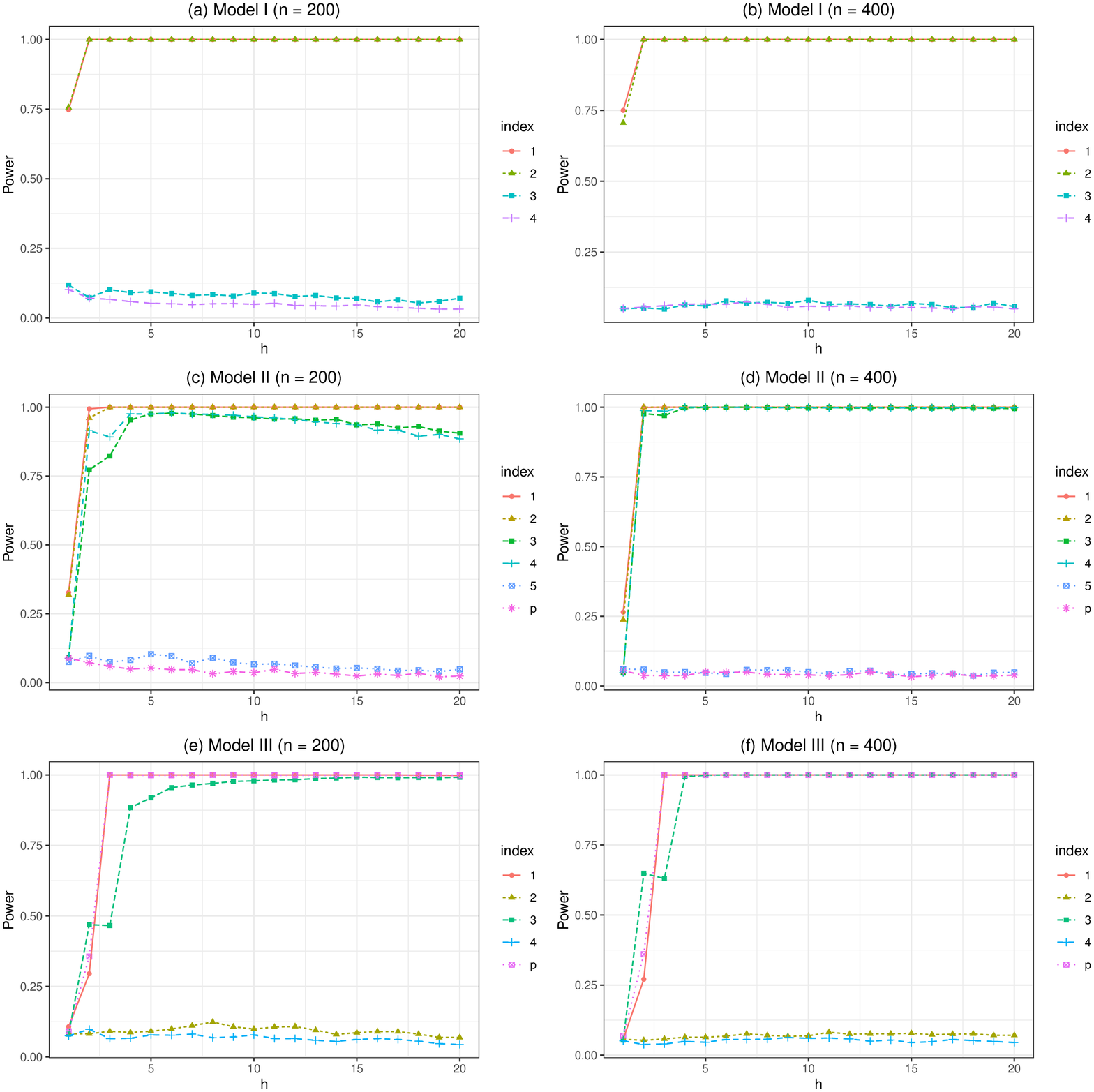}
    \caption{The plot of empirical power  with respect to $h$ when the Lasso penalty is used.}
    \label{fig1}
\end{figure}

Figure \ref{fig1} presents the empirical power functions with
respect to $h$ for several $j$s when the penalty in the penalized
least squares is taken to be the Lasso penalty. The results for
the SCAD penalty are similar to those for the Lasso penalty, and
are depicted in Figure~\ref{figs1} in the supplement. As shown in
Figure~\ref{fig1}, it can be seen that for truly active variables,
the empirical power has a sharp increase as $h$ increases if
$h\leqslant d+1$, and remains stable in a wide range of $h$. It is
also observed that the empirical power decreases slowly as $h$
increases when $h$ is much larger than $d$ and $n=200$. It can
also be seen from Figure~\ref{fig1} that power of some active
variables in Model II does not approach to 1. In particular, the
empirical power of $X_3$ in Model II achieves its maximum around
$h=5$, but its value is slightly less than $1$. This phenomenon
disappears when the sample size increases from 200 to 400.
Figure~\ref{fig1} also implies that the empirical powers stand
firm around $1$ when $h>d+1$ for $n = 400$. For inactive
variables, the empirical size always stays around nominal level
$\alpha = 0.05$. Figure~\ref{fig1} and \ref{figs1} seems to imply that
$h=5$ is good choice for Models I, II and III. Thus, we set $h=5$
in Sections 5.2, 5.3 and 5.4.



\subsection{Comparison with existing methods}

This subsection is devoted to comparing the performance of the
proposed procedure with two popular methods: the LDPE method
\citep{zhang2014confidence} and the decorrelated score method
\citep{ning2017general} by pretending all the observations are
characterized by a linear model. Let $T^{ZZ}$ and $T^{NL}$ stands for
these two test procedures, respectively. We compare the
performance of the three procedures in terms of empirical size and power. The
proposed test uses linear B-spline transformation with $h=5$.

The data were generated from Models I, II and III described in
Section 5.1. We summarize the simulation results in terms of rejection
rate, which is defined to be the proportion of $p$-values being
smaller than the nominal level $\alpha=0.05$ based on 1000
replications. In each simulation, we test $H_{0j}$ for
$j=1,\cdots,p$. For inactive predictors (i.e., $X_j$ with $j\in
{\cal A}^c$), it is expected that the proportion is close to the
nominal level, while for active predictors, it is expected to
reject $H_{0j}$ with a large probability, and the rejection rate
is the estimated power of the underlying test.

\begin{table}[htbp!]
\centering
\caption{Empirical rejection rate of $H_{0j}$ with $\alpha=5\%$ when the Lasso penalty is used. }
\label{tab1}
\begin{tabular}{@{}cccccccccccc@{}}
\toprule
$n$ & Method         & $X_1$ & $X_2$ & $X_3$ & $X_4$ & $X_5$ & $X_{1996}$ & $X_{1997}$ & $X_{1998}$ & $X_{1999}$ & $X_{2000}$ \\
\hline
\multicolumn{12}{c}{Model I: $X_1$ and $X_2$ are active predictors }\\
\hline
200 & $W_n$-Lasso    & 1.000 & 1.000 & 0.077 & 0.053 & 0.058 & 0.038      & 0.052      & 0.052      & 0.047      & 0.047      \\
 & $T^{NL}$-Lasso  & 1.000 & 1.000 & 0.057 & 0.040 & 0.032 & 0.030      & 0.030      & 0.032      & 0.037      & 0.040     \\
 & $T^{ZZ}$        & 1.000 & 1.000 & 0.141 & 0.070 & 0.054 & 0.043      & 0.049      & 0.042      & 0.053      & 0.056      \\
\hline
400 & $W_n$-Lasso     & 1.000 & 1.000 & 0.059 & 0.067 & 0.048 & 0.052      & 0.046      & 0.042      & 0.048      & 0.059      \\
 & $T^{NL}$-Lasso  & 1.000 & 1.000 & 0.045 & 0.039 & 0.043 & 0.041      & 0.039      & 0.033      & 0.035      & 0.042      \\
 & $T^{ZZ}$        & 1.000 & 1.000 & 0.093 & 0.063 & 0.060 & 0.056      & 0.055      & 0.048      & 0.043      & 0.052      \\ \bottomrule
\hline
\multicolumn{12}{c}{Model II: $X_1, X_2, X_3$ and $X_4$ are active predictors}\\
\hline
200 & $W_n$-Lasso     & 1.000                         & 1.000 & 0.976 & 0.975 & 0.074 & 0.058      & 0.036      & 0.041      & 0.043      & 0.053      \\
 & $T^{NL}$-Lasso  & 1.000                         & 0.999 & 0.027 & 0.046 & 0.027 & 0.031      & 0.032      & 0.017      & 0.039      & 0.025      \\
 & $T^{ZZ}$        & 1.000                         & 1.000 & 0.041 & 0.038 & 0.045 & 0.049      & 0.055      & 0.042      & 0.057      & 0.039      \\
\hline
400 & $W_n$-Lasso      & 1.000                         & 1.000 & 0.999 & 1.000 & 0.047 & 0.047      & 0.057      & 0.037      & 0.055      & 0.046   \\
 & $T^{NL}$-Lasso  & 1.000                         & 1.000 & 0.064 & 0.101 & 0.037 & 0.033      & 0.035      & 0.034      & 0.034      & 0.031      \\
 & $T^{ZZ}$       & 1.000                         & 1.000 & 0.029 & 0.082 & 0.040 & 0.047      & 0.048      & 0.047      & 0.045      & 0.046     \\
\bottomrule\hline
\multicolumn{12}{c}{Model III: $X_1, X_3$ and $X_p$ are active predictors}\\
\hline
200 & $W_n$-Lasso      & 1.000 & 0.074 & 0.919 & 0.078 & 0.059 & 0.043      & 0.041      & 0.051      & 0.080      & 0.999      \\
 & $T^{NL}$-Lasso  & 0.240 & 0.051 & 0.440 & 0.049 & 0.032 & 0.027      & 0.015      & 0.035      & 0.036      & 0.255      \\
 & $T^{ZZ}$        & 0.285 & 0.078 & 0.499 & 0.083 & 0.052 & 0.046      & 0.034      & 0.055      & 0.052      & 0.294      \\
\hline
400 & $W_n$-Lasso     & 1.000 & 0.066 & 0.999 & 0.042 & 0.044 & 0.048      & 0.029      & 0.058      & 0.066      & 1.000      \\
 & $T^{NL}$-Lasso  & 0.350 & 0.049 & 0.461 & 0.053 & 0.032 & 0.040      & 0.025      & 0.040      & 0.037      & 0.329      \\
 & $T^{ZZ}$    & 0.358 & 0.063 & 0.496 & 0.064 & 0.042 & 0.054      & 0.040      & 0.047      & 0.050      & 0.343      \\
\bottomrule\hline
\end{tabular}
\end{table}

The results are summarized in Tables \ref{tab1} and \ref{tabs1} when the Lasso penalty and the SCAD penalty are used, respectively.
These two tables only display the rejection rates for predictors $X_1$-$X_5$ and $X_{1996}$-$X_{2000}$
to save space. From these two tables, it can be seen that for linear model (Model I),
the newly proposed test performs as well as the other two tests.
All three tests control the size well and have good empirical power.
For nonlinear model (Model II) and heteroscedastic error model (Model III),
all three tests also control Type I error at the desired nominal level.
But the newly proposed test has much higher power than the other two tests.
In particular, the power of the proposed test statistic $W_{nj}$ for Model II is very close to $1$
under $H_{03}: 3\in\mathcal{A}^c$, while the empirical powers of $T^{ZZ}$ and $T^{NL}$
are around 0.05 even when $n=400$.


\subsection{Comparison of FDR control procedures}

We next investigate the finite sample performance of the proposed FDR control procedures
by simultaneously testing $H_{0 j}: j \in \mathcal{A}^{c}$ for all $j=1, \cdots, p$.
This simulation study is designed to
compare the newly proposed FDR control procedure with
model-X knockoff procedures \citep{candes2018panning}, for which we consider the Lasso coefficient-difference (LCD) statistic.

To understand the impact of sparsity level on FDR control procedures, we consider the following two models, which are modified from
Models I and II.

Model IV: $Y  = \boldsymbol{X}^{\top}\boldsymbol{\beta}_1 + \epsilon.$

Model V: $Y =(\boldsymbol{X}^{\top}\boldsymbol{\beta}_2)/\{0.5+\left(1.5+X_{p-1} + X_{p-2}\right)^{2}\}+0.1\epsilon.$

In our simulation, we set $\boldsymbol{\beta}_1= (\overbrace{1,\cdots,1}^{s_1},0,\cdots,0)$ and
$\boldsymbol{\beta}_2 = (\overbrace{1,\cdots,1}^{s_2},0,\cdots,0)$.
The sparsity levels of Model IV and V are $s_1$ and $s_2 + 2$, respectively.
The sparsity levels of each model are set to be 4, 6, and 8. The distribution of $\boldsymbol{X}$ and $\epsilon$
are the same as those for Models I and II.
In this simulation study, $p=2000$ and the sample size $n=200$ and 400.
For the newly proposed FDR control procedure, the Lasso is also used, and we set
$h=5$. We search the $\hat{t}$ over $ [0, 2\log p + 0.75\log \log p]$ and set
$\hat{t} = 2\log p + 4\log \log p$ if $\hat{t}$ does not exist over the interval.

The empirical FDR and empirical power over 1000 replications are presented in Table \ref{tab:FDR} at
the significance level $\alpha=0.1$ or $0.2$, where
$\mbox{Power}={|\widehat{\mathcal{A}} \cap \mathcal{A}|}/{|\mathcal{A}|}$
with $\widehat{\mathcal{A}}=\left\{j: W_{nj} \geqslant \hat{t}\right\}$.
From Table~\ref{tab:FDR}, it can be seen that
%
for Models VI and V, the proposed procedure and the model-X knockoff procedures can control the FDR well.
For nonlinear model (model V), the proposed method generally has larger powers than the model-X knockoff procedures.
The Knockoff+, a more conservative knockoff procedure, has much conservative FDR
and therefore has very low power under the setting of Model V.

\begin{singlespace}
\begin{table}[htbp!]
    \centering
    \caption{Empirical size and power of FDR control procedures
    at nominal level $\alpha = 0.1,\ 0.2$.
    }
    \label{tab:FDR}
    \begin{tabular}{cc|cccc|cccc}
        \hline
        &               & \multicolumn{4}{c|}{Model IV}  & \multicolumn{4}{c}{Model V} \\ \hline
        &       & \multicolumn{2}{c}{$\alpha=0.1$}     & \multicolumn{2}{c|}{$\alpha=0.2$}
        & \multicolumn{2}{c}{$\alpha=0.1$}      & \multicolumn{2}{c}{$\alpha=0.2$}      \\ \hline
        Sparsity            & Method        & FDR          & Power        & FDR          & Power        & FDR           & Power        & FDR           & Power        \\ \hline
        \multicolumn{10}{c}{$n=200$}\\
        \hline
        \multirow{3}{*}{4}  & Proposed      & 0.089        & 0.954        & 0.104        & 0.954        & 0.086         & 0.954        & 0.118         & 0.979        \\ 
        & LCD-Knockoff  & 0.061        & 1.000        & 0.154        & 1.000        & 0.062         & 0.500        & 0.129         & 0.500        \\ 
        & LCD-Knockoff+ & 0.008        & 0.012        & 0.153        & 0.480        & 0.002         & 0.002        & 0.065         & 0.049        \\ \hline
        \multirow{3}{*}{6}  & Proposed      & 0.083        & 0.787        & 0.123        & 0.800        & 0.089         & 0.860        & 0.148         & 0.869        \\ 
        & LCD-Knockoff  & 0.065        & 1.000        & 0.155        & 1.000        & 0.063         & 0.638        & 0.158         & 0.646        \\ 
        & LCD-Knockoff+ & 0.023        & 0.051        & 0.137        & 1.000        & 0.007         & 0.007        & 0.152         & 0.314        \\ \hline
        \multirow{3}{*}{8}  & Proposed      & 0.097        & 0.600        & 0.151        & 0.654        & 0.087         & 0.693        & 0.163         & 0.745        \\ 
        & LCD-Knockoff  & 0.080        & 1.000        & 0.164        & 1.000        & 0.068         & 0.628        & 0.160         & 0.664        \\ 
        & LCD-Knockoff+ & 0.063        & 0.236        & 0.158        & 1.000        & 0.017         & 0.026        & 0.151         & 0.548        \\ 
        \bottomrule\hline
        \multicolumn{10}{c}{$n=400$}\\
        \hline
        \multirow{3}{*}{4}  & Proposed      & 0.083        & 1.000        & 0.094        & 1.000        & 0.064         & 0.970        & 0.110         & 0.970        \\ 
        & LCD-Knockoff  & 0.061        & 1.000        & 0.151        & 1.000        & 0.060      & 0.500      & 0.133      & 0.500     \\ 
        & LCD-Knockoff+ & 0.016        & 0.026        & 0.153        & 0.482        & 0             & 0            & 0.059      & 0.046     \\
\hline
        \multirow{3}{*}{6}  & Proposed      & 0.066        & 0.999        & 0.139        & 0.999        & 0.076         & 0.992        & 0.152         & 0.993        \\ 
        & LCD-Knockoff  & 0.067        & 1.000        & 0.151        & 1.000        & 0.058         & 0.667        & 0.151         & 0.667        \\ 
        & LCD-Knockoff+ & 0.026        & 0.060        & 0.141        & 1.000        & 0.004         & 0.004        & 0.147         & 0.321        \\ \hline 
        \multirow{3}{*}{8}  & Proposed      & 0.056        & 0.987        & 0.173        & 0.994        & 0.064         & 0.987        & 0.183         & 0.993        \\ 
        & LCD-Knockoff  & 0.078        & 1.000        & 0.158        & 1.000        & 0.061         & 0.739        & 0.151         & 0.744        \\ 
        & LCD-Knockoff+ & 0.065        & 0.257        & 0.154        & 1.000        & 0.016         & 0.028        & 0.132         & 0.735        \\ \hline 
    \end{tabular}
\end{table}
\end{singlespace}

\subsection{A real data example}

We illustrate the proposed procedures by an empirical analysis of the data studied in
\cite{scheetz2006regulation}, \cite{huang2010variable} and \cite{li2017variable}.
This data set contains 120 twelve-week old male rats from which over 31042 different probes (genes) from eye tissue were measured. The intensity values were normalized using the RMA (robust multi-chip averaging, \cite{bolstad2003comparison}) method to obtain summary expression values for each probe. We are interested in finding the probes that could significantly affect the gene TRIM32. This gene was reported in \cite{chiang2006homozygosity} and was believed to cause Bandet-Biedl syndrome which is a genetically heterogeneous disease of multiple organ systems including the retina.

Based on the procedure in \cite{scheetz2006regulation}, we first exclude probes that were not expressed in the eye or that lacked sufficient variation. As a result, a total of 18976 probes were considered ``sufficient variables". Among these 18976 probes, we further apply the procedure reported in \cite{huang2008adaptive} to select 3000 probes with large variances. It is expected that only a few genes are related to TRIM32.

We apply the newly proposed procedures to test whether the $j$th gene
is associated with TRIM32.
The proposed FDR control procedure identifies 19 probes at the nominal level 0.05.
The IDs of identified probes and their p-values are listed in Table \ref{tab6}.

Among these 19 probes, 1383110\_at is also identified by \cite{huang2008adaptive,huang2010variable,wang2012quantile,li2017variable}. Other probes such as 1379597\_at, 1382263\_at and 1390401\_at are repeatedly detected by other selection methods \citep{wang2012quantile,huang2008adaptive}. Due to the complexity of gene microarray data, it is not surprised that different approaches
detect different important gene sets. Based on these repeated findings, we are more confident that the probes selected by the proposed procedures provide useful information for further biological study.

\begin{table}[htbp]
    \centering
    \caption{Probe ID selected.}
    \label{tab6}
    \begin{tabular}{cc|cc}
        \hline
        Probe ID         & P-value & Probe ID         & P-value \\
        \hline
        1383110\_at      & $3.04\times 10^{-6}$ &  1383983\_at      & $8.28\times 10^{-5}$ \\
        1379597\_at      & $1.17\times 10^{-5}$ & 1393021\_at      & $8.86\times 10^{-5}$ \\
        1382263\_at      & $1.30\times 10^{-5}$ & 1371081\_at      & $1.19\times 10^{-4}$ \\
        1380978\_at      & $2.30\times 10^{-5}$ & 1370551\_a\_at   & $1.24\times 10^{-4}$ \\
        1389460\_at      & $2.34\times 10^{-5}$ & 1382517\_at      & $1.39\times 10^{-4}$ \\
        1390401\_at      & $3.34\times 10^{-5}$ & 1393684\_at      & $1.41\times 10^{-4}$ \\
        1369453\_at      & $4.07\times 10^{-5}$ & 1381515\_at      & $1.56\times 10^{-4}$ \\
        1376829\_at      & $4.21\times 10^{-5}$ & 1389955\_at      & $1.59\times 10^{-4}$ \\
        1379818\_at      & $5.32\times 10^{-5}$ & 1373177\_x\_at   & $1.65\times 10^{-4}$ \\
        1382743\_at      & $5.94\times 10^{-5}$ &  \\
        \bottomrule\hline
    \end{tabular}
\end{table}

\section{Conclusions and discussions}
In this paper, we have developed model-free inference procedures
for high-dimensional data by using a new reformation within sufficient dimension reduction framework
and the orthogonalization technique. The proposed test statistic is shown to asymptotically follow
$\chi^2$ distribution under the null hypothesis, and a non-central $\chi^2$ distribution under local alternative hypotheses.
The FDR control for a large-scale multiple testing problem
using the proposed test to identify important predictors is also studied.
We introduce the maximum correlation coefficient to quantify the dependence, and
develop a truncated BH method to control the FDR. We show that the procedure is valid for controlling FDR
under some mild conditions.

\section*{Acknowledgment}
X. Guo's research was supported by National Natural Science Foundation of China Grant No. 12071038.
R. Li and Z. Zhang's research was supported by a National Science Foundation grant
DMS 1820702. C. Zou's research was supported by NNSF of China Grants (Nos. 11925106, 11931001 and 11971247)
and NSF of Tianjin Grant (No. 18JCJQJC46000).

\section*{Appendix}

\setcounter{equation}{0}
\renewcommand{\theequation}{A.\arabic{equation}}

\noindent{\bf A.1 Proof of Theorem 1}.
Let us start the proof of Part (a). Note that
$\bar{S}_{nk}^j=S_{nk}^{(1)}+S_{nk}^{(2)}$, where
\begin{equation*}
S_{nk}^{(1)}=\frac{1}{\sqrt n}\sum_{i=1}^n\eta_{ik}^j(X_{ij}-\bZ^{\top}_{ij}\hat{\btheta}_j)
\quad\text{and}\quad S_{nk}^{(2)}=\frac{(\bgamma_{kj}-\hat{\bgamma}_{kj})^{\top}}{\sqrt n}\sum_{i=1}^n\bZ_{ij}(X_{ij}-\bZ^{\top}_{ij}\hat{\btheta}_j).
\end{equation*}
We first show that $S_{nk}^{(2)}$ is asymptotically negligible.
Note that
\[
|S_{nk}^{(2)}|\leqslant \sqrt n\|\hat{\bgamma}_{kj}-\bgamma_{kj}\|_{1}
\|\frac{1}{n}\sum_{i=1}^n\bZ_{ij}(X_{ij}-\bZ^{\top}_{ij}\hat{\boldsymbol{\theta}}_j)\|_{\infty}.
\]
It follows by Condition (B1) that
$
\|\hat{\bgamma}_{kj}-\bgamma_{kj}\|_{1}=O_p(\lambda_Ys_Y).
$
Furthermore, it can be shown that
$\|\frac{1}{n}\sum_{i=1}^n\bZ_{ij}(X_{ij}-\bZ^{\top}_{ij}\hat{\btheta}_j)\|_{\infty}\leqslant C\lambda_X
$
by using the Karush-Kuhn-Tucker condition and similar arguments to those in \cite{loh2015regularized}.
Thus  it follows that
$|S_{nk}^{(2)}|
=O_p(\sqrt n\lambda_Y\lambda_Xs_Y)=o_p(1)
$
since  $s_Y=o(\sqrt n/\log p)$ and $\lambda_X, \lambda_Y = O(\sqrt{\frac{\log p}{n}})$ in Condition (B3).

We next deal with $S_{nk}^{(1)}$. Note that $S_{nk}^{(1)}=S_{11}+S_{12}$, where
\[
S_{11}=\frac{1}{\sqrt n}\sum_{i=1}^n\eta_{ik}^j(X_{ij}-\bZ^{\top}_{ij}\boldsymbol{\theta}_j^*),
\quad\text{and}\ S_{12}=
\frac{(\boldsymbol{\theta}^*_j-\hat{\boldsymbol{\theta}}_j)^{\top}}{\sqrt n}\sum_{i=1}^n\eta_{ik}^j\bZ_{ij}.
\]
It can be shown by using the central limit theorem that $S_{11}$ converges to a normal distribution.
Thus, we focus on $S_{12}$.
Let $Z_{j,l}$ be the $l$th element of $\bZ_{j}$. Under Condition (B2) and using
Lemma 1.7.2 in \cite{vershynin2018high}, $\eta_k^jZ_{j,l}$ is sub-exponential.
Then it follows from Corollary 2.8.3 in \cite{vershynin2018high} that
$$
\Pr(|\frac{1}{n}\sum_{i=1}^n\eta_{ik}^jZ_{ij,l}|\geqslant t)\leqslant 2\exp
\left[-Cn\min(\frac{t^2}{K_{l}^2},\frac{t}{K_{l}})\right],
$$
where $K_{l}=\|\eta_{k}Z_{j,l}\|_{\varphi_1}$. Then it follows that
\begin{align*}
&\Pr\left(\frac{1}{n}\|\sum_{i=1}^n\eta_{ik}^j\bZ_{ij}\|_{\infty}>t\right)
\leqslant 2p\max_{1\le l\le p}\exp
\left[-Cn\min(\frac{t^2}{K_{l}^2},\frac{t}{K_{l}})\right]\leqslant 2p\exp
\left[-Cn\min(\frac{t^2}{K^2},\frac{t}{K})\right],
\end{align*}
where $K=\max\limits_{l} K_{l}$, which is finite by Condition (B2).

Setting $t=C'\sqrt{\log p/n}$ and $C'$ be a large enough constant, it leads to
\begin{eqnarray*}
\|\frac{1}{\sqrt n}\sum_{i=1}^n\eta_{ik}^j\bZ_{ij}\|_{\infty}=O_p(\sqrt{\log p}).
\end{eqnarray*}
On the other hand, under Condition (B1),
$\|\boldsymbol{\theta}^*_j-\hat{\boldsymbol{\theta}}_j\|_1=O_p(\lambda_Xs_X).
$
Thus when $s_X=o(\sqrt n/\log p)$, it follows that $
|S_{12}|\leqslant \|\frac{1}{\sqrt n}\sum_{i=1}^n\eta_{ik}^j\bZ_{ij}\|_{\infty}\|\boldsymbol{\theta}^*_j-\hat{\boldsymbol{\theta}}_j\|_1=O_p(\lambda_Xs_X\sqrt{\log p})=o_p(1).
$

In summary, $\bar{S}_{nk}^j=S_{11}+o_p(1)=\frac{1}{\sqrt n}\sum_{i=1}^n\eta_{ik}^j(X_{ij}-\bZ^{\top}_{ij}\boldsymbol{\theta}^*_j)+o_p(1).$
Denote $\boldsymbol{\Omega}_j$ to be a $h\times h$ matrix with $(l, k)$-element being
$\mbE[\eta_l^j\eta_k^j(X_{j}-\bZ^{\top}_{j}\boldsymbol{\theta}^*_j)^2]$. By definition of $\bar{\boldsymbol{S}}_n^j$, it follows that
$\bar{\boldsymbol{S}}_n^j \rightarrow N_h(0, \boldsymbol{\Omega}_j).$

As a result, we have that
$\bar{\boldsymbol{S}}_n^{j\top}\boldsymbol{\Omega}^{-1}_j\bar{\boldsymbol{S}}_n^j\rightarrow \chi^2_h.$
In practice, $\boldsymbol{\Omega}_j$ is generally unknown,
but can be consistently estimated by $\hat{\boldsymbol{\Omega}}_j =
\frac{1}{n}\sum_{i=1}^n \boldsymbol{S}_i^j\boldsymbol{S}_i^{j{\top}}$, whose $(l, k)$-element is
\begin{align*}
&\widehat{\omega}_{l k}=\frac{1}{n}\sum_{i=1}^n(f_l(Y_i)-\bZ^{\top}_{ij}\hat\gamma_l)(f_k(Y_i)-\bZ^{\top}_{ij}\hat{\bgamma}_{kj})
(X_{ij}-\bZ^{\top}_{ij}\hat{\boldsymbol{\theta}}_j)^2.
\end{align*}
The consistency of $\hat{\boldsymbol{\Omega}}_j$ to $\boldsymbol{\Omega}_j$ will follow
from the consistency of $\bZ^{\top}_{ij}\hat\bgamma_l, \bZ^{\top}_{ij}\hat{\bgamma}_{kj}, \bZ^\top_{ij}\hat{\boldsymbol{\theta}}_j$ to $\bZ^{\top}_{ij}\boldsymbol{\gamma}_l, \bZ^{\top}_{ij}\bgamma_{kj}, \bZ^{\top}_{ij}\boldsymbol{\theta}^*_j$, respectively.
Thus, it follows that $\max_{1\le i\le n} |\bZ^{\top}_{ij}(\hat{\boldsymbol{\theta}}_j
-\boldsymbol{\theta}^*_j)|=o_p(1)$. Other consistency results can be established by
similar arguments.
Note that
\begin{align*}
&\bZ^{\top}_{ij}(\hat{\boldsymbol{\theta}}_j-\boldsymbol{\theta}^*_j)=(\bZ_{ij}-\mu_{\bZ_j})^{\top}
(\hat{\boldsymbol{\theta}}_j-\boldsymbol{\theta}^*_j)+\mu_{\bZ_j}^{\top}(\hat{\boldsymbol{\theta}}_j-\boldsymbol{\theta}^*_j);\\
&|\mu_{\bZ_j}^{\top}(\hat{\boldsymbol{\theta}}_j-\boldsymbol{\theta}^*_j)|\leqslant \|\mu_{\bZ_j}\|_{\infty}\|\hat{\boldsymbol{\theta}}_j-\boldsymbol{\theta}^*_j\|_1=o_p(1).
\end{align*}
To control the term $(\bZ_{ij}-\mu_{\bZ_j})^{\top}(\hat{\boldsymbol{\theta}}_j-\boldsymbol{\theta}^*_j)$,
we first derive the order of the term $\max_{1\le i\le n} \|\bZ_{ij}-\mu_{\bZ_j}\|_{\infty}$.
Under Condition (B2), it follows that
\begin{align*}
\Pr(\max_{1\le i\le n} \|\bZ_{ij}-\mu_{\bZ_j}\|_{\infty}>t)
\leqslant 2np\max_{1\le l\le p}\exp\left(-C\frac{t^2}{J_l}\right)\leqslant 2np\exp\left(-C\frac{t^2}{J}\right),
\end{align*}
where $J=\max_l J_l, J_l=\|Z_{j,l}-\mu_{Z_{j,l}}\|^2_{\varphi_2}$. Letting $t=C''\sqrt{\log n+\log p}$
with $C''$ being a large constant, it follows that
\begin{eqnarray*}
\max_{1\le i\le n} \|\bZ_{ij}-\mu_{\bZ_j}\|_{\infty}=O_p(\sqrt{\log n+\log p}).
\end{eqnarray*}
Thus, $\max_{1\le i\le n} |\bZ^{\top}_{ij}(\hat{\boldsymbol{\theta}}_j-\boldsymbol{\theta}^*_j)|=o_p(1).$
Therefore, under condition both $s_{X}$ and  $s_{Y}$ is of order $o(\sqrt{n} / \log p),$
$
W_{nj}=\bar{\boldsymbol{S}}^{j\top}_n\hat{\boldsymbol{\Omega}}_j^{-1}\bar{\boldsymbol{S}}_n^j\rightarrow \chi^2_h
$
in distribution.

We next prove Part (b).  Consider $\bar{S}^j_{n k}$. Similar to the proof of Part (a), we have
$$
\bar{S}_{n k}^j=\frac{1}{\sqrt{n}} \sum_{i=1}^{n} \eta_{ik}^j\left(X_{i j}-\bZ_{i j}^{\top}
 \boldsymbol{\theta}^*_j\right)+o_{p}(1),
$$
where $\eta_{k}^j=f_{k}(Y)-\bZ_{j}^{\top}\bgamma_{kj}, \bgamma_{kj}=\mbE\left[\bZ_{j} \bZ_{j}^{\top}\right]^{-1} \mbE\left[\bZ_{j} f_{k}(Y)\right]$ and $\boldsymbol{\theta}^*_j=\mbE\left[\bZ_{j} \bZ_{j}^{\top}\right]^{-1} \mbE\left[\bZ_{j} X_{j}\right] .$ Under local
alternative hypotheses $H_{1 n}, f_{k}(Y)=\boldsymbol{X}^{\top}\boldsymbol{\beta}_{k}^{0}+\epsilon_{k},$
with $E\left[\epsilon_{k} \boldsymbol{X}\right]=0 .$ Thus,
$$
\eta_{k}^j=\epsilon_{k}+X_{j} \beta_{j k}^{0}-\beta_{j k}^{0} \boldsymbol{\theta}^{* \top}_j \bZ_{j}.
$$
Since $\mbE\left[\btheta^{* \top}_j \bZ_{j}\left(X_{j}-\bZ_{j}^{\top} \boldsymbol{\theta}^*_j\right)\right]=0$,
we have
$$
\begin{aligned}
\bar{S}_{n k}^j&=\frac{1}{\sqrt{n}} \sum_{i=1}^{n} \epsilon_{i k}\left(X_{i j}  -\bZ_{i j}^{\top} \boldsymbol{\theta}^*_j\right)+\frac{\sqrt{n}
\beta_{j k}^{0}}{n} \sum_{i=1}^{n} X_{i j}\left(X_{i j}-\bZ_{i j}^{\top}\boldsymbol{\theta}^*_j\right)\\
&\quad -\frac{\sqrt{n} \beta_{j k}^{0}}{n} \sum_{i=1}^{n} \btheta^{* \top}_j \bZ_{i j}\left(X_{i j}-\bZ_{i j}^{\top}\boldsymbol{\theta}^*_j\right)+o_{p}(1) \\
&=\frac{1}{\sqrt{n}} \sum_{i=1}^{n} \epsilon_{i k}\left(X_{i j}-\bZ_{i j}^{\top}
\boldsymbol{\theta}^*_j\right)+\frac{\sqrt{n} \beta_{j k}^{0}}{n} \sum_{i=1}^{n} X_{i j}\left(X_{i j}-\bZ_{i j}^{\top} \boldsymbol{\theta}^*_j\right)+o_{p}(1).
\end{aligned}
$$
By definition of $\bar{\boldsymbol{S}}_{n}^j$, we have
$\bar{\boldsymbol{S}}_{n}^j \rightarrow N_{h}(\delta_j\boldsymbol{b} ,\boldsymbol{\Omega}_j),$
where $\delta_j=\mbE(X_{j}^{2})-\mbE(X_{j} \bZ_{j}^{\top}) \mbE(\bZ_{j}
\bZ_{j}^{\top})^{-1} \mbE(\bZ_{j} X_{j})$.
Thus,  $W_{nj}\to \chi_h^2(\delta_j^2 \boldsymbol{b}^{\top}\boldsymbol{\Omega}_j^{-1}\boldsymbol{b})$. This
completes the proof of Theorem 1.

\bigskip

\noindent{\bf A.2 Proof of Theorem 2}.
Set $b_p = 2\log p + 2d_0\log\log p$ and $a_p = 2\log p+ (h-1)\log \log p$. If $\hat{t}$ does not exist in $[0,b_p]$, we consider the event $J_0 = \left\{ \sum\limits_{j \in \mathcal{A}^c} I(W_{nj}\geqslant a_p)\geqslant 1\right\}$. By Proposition \ref{prop1}, it holds
\begin{align*}
    &\quad\ \Pr(\sum_{j\in\mathcal{A}^c}I(W_{nj}\geqslant a_p)\geqslant 1)=\Pr(\exists j\in\mathcal{A}^c, \widetilde W_{nj}+W_{nj}-\widetilde W_{nj}\geqslant a_p)\\
    &\leqslant \Pr(\exists j\in\mathcal{A}^c, \widetilde W_{nj}\geqslant a_p-Co(1))+\Pr(\exists j\in\mathcal{A}^c, W_{nj}-\widetilde W_{nj}\geqslant Co(1))\\
    &\leqslant p\max\limits_{j\in \mA^c}\Pr(\widetilde W_{nj}\geqslant a_p)+o(1)\\
    &=p\max\limits_{j\in \mA^c}\Pr(\chi^2_h\geqslant a_p)(1+o(1))+o(1)\\
    &=O(\frac{1}{\sqrt{\log p}}) + o(1) =o(1).
\end{align*}
The second to last equality follows by Theorem 3.1 and Lemma 3.2 in \cite{cai2019large}
which implies that $\max\limits_{j \in \mA^c}\left| \frac{\Pr(\widetilde W_{nj} \geqslant t)}{G(t)}
 - 1\right| = O\left((\log p)^{-3/2}\right)$ uniformly in $t\in [0,a_p]$.
 The last equality comes from (\ref{tail2: 1}) in Lemma \ref{tailbound}. Hence, $\mbox{FDP}(a_p)\to 0$ with probability 1.

Now consider the case when $0 \leq \hat{t} \leq b_{p}$ holds. We have
$$
\operatorname{FDP}(\hat{t})=\frac{\sum_{j \in \mathcal{A}^c} I\left\{W_{nj}\geq \hat{t}\right\}}{\max \left\{\sum_{j=1}^{p} I\left\{W_{nj}\geq \hat{t}\right\}, 1\right\}} \leq \frac{|\mathcal{A}^c| G(\hat{t})}{\max \left\{\sum_{j=1}^{p} I\left\{W_{nj}\geq \hat{t}\right\}, 1\right\}}\left(1+A_{p}\right),
$$
where $A_{p}=\sup _{0 \leq t \leq b_{p}}\left|\frac{\sum_{j \in \mathcal{A}^c} I\left\{W_{nj}\geq t\right\}}{|\mathcal{A}^c| G(t)}-1\right|$. Note that by definition of $\hat t$, $\frac{|\mathcal{A}^c|G(\hat{t})}{\max \bigl\{\sum_{j=1}^{p} I\{W_{nj} \geq \hat{t}\}, 1\bigr\}} \leq \alpha$. The proof is complete if $A_{p} \rightarrow 0$ in probability.

We next show
\begin{equation}
    \sup _{0 \leq t \leq b_p}\left|\frac{\sum_{j \in \mathcal{A}^{c}} I\left\{W_{nj} \geq t\right\}}{|\mathcal{A}^c|G(t)}-1\right| \rightarrow 0.
    \label{ea.10}
\end{equation}
Recall the Condition D2 that $\left|\widetilde{W}_{nj} - W_{nj}\right| = o_p(1)$ uniformly for all $j$, based on Lemma \ref{tailrate} we have $G(t+o(1))/G(t) = 1 + o(1)$ uniformly for $ t\in [0, b_p]$. Thus, to show (\ref{ea.10}),
it is sufficient to show that
$$
\sup _{0 \leqslant t \leqslant b_p}\left|\frac{\sum\limits_{j \in \mathcal{A}^{c}} I\left\{\widetilde{W}_{nj} \geqslant t\right\}}{|\mathcal{A}^c|G(t)}-1\right| \rightarrow 0.
$$
To this aim, it suffices to show that
$$
\sup _{0 \leqslant t \leqslant b_p}\left|\frac{\sum\limits_{j \in \mathcal{A}^{c}}\left\{I\left(\widetilde{W}_{nj} \geqslant t\right)-G(t)\right\}}{p G(t)}\right| \rightarrow 0
$$
in probability since $|\mA^c|/p\rightarrow 1$. Define $v_p = {1}/{\log\log p}$ and $t_l = lv_p$, for
$l = 1,\cdots,L-1$, and $t_0 =  0$, $t_L = b_p$. Thus for any $t\in [0,b_p]$, there exists
$l\in \left\{1,\cdots,L-1\right\}$ such that $t_{l}\leqslant t\leqslant t_{l+1}$. Then we have
$$
\frac{\sum\limits_{j \in \mathcal{A}^{c}} I\left(\left|\widetilde{W}_{nj}\right| \geqslant t_{l+1}\right)}{p G\left(t_{l+1}
\right)} \frac{G\left(t_{l+1}\right)}{G\left(t_{l}\right)}
\leqslant \frac{\sum\limits_{j \in \mathcal{A}^{c}} I\left(\left|\widetilde{W}_{nj}\right|
\geqslant t\right)}{p G\left(t\right)} \leqslant \frac{\sum\limits_{j \in \mathcal{A}^{c}} I\left(\left|\widetilde{W}_{nj}\right|
\geqslant t_{l}\right)}{p G\left(t_{l}\right)} \frac{G\left(t_{l}\right)}{G\left(t_{l+1}\right)}  .
$$
Since $\frac{G(t_{l+1})}{G(t_l)} \rightarrow 1$ uniformly over $l \in \left\{1,\cdots,L-1\right\}$, according to the Lemma 6.3 in \cite{liu2013gaussian}, it is sufficient to show that
\begin{eqnarray}
&&\int_{0}^{b_{p}} \Pr\left[\left|\frac{\sum\limits_{j \in \mathcal{A}^{c}}
\left\{I\left(\left|\widetilde{W}_{nj}\right| \geqslant t\right)-\Pr\left(\widetilde{W}_{nj} \geqslant t\right)\right\}}{pG(t)}\right| \geqslant \epsilon\right] \mathrm{d} t=o(v_{p})
\label{ea.11}\\
&&\sum_{l=L-1}^{L} \Pr\left[\left|\frac{\sum\limits_{j \in \mathcal{A}^{c}}\left\{I\left(\widetilde{W}_{nj} \geqslant t_{l}\right)-G\left(t_{l}\right)\right\}}{p G\left(t_{l}\right)}\right| \geqslant \epsilon\right]  = o(1).
\label{ea.12}
\end{eqnarray}
We will show only (\ref{ea.11}) since (\ref{ea.12}) can be proved in a similar way. We next show (\ref{ea.11}).
By Markov inequality, it is enough to show that
\[
\int_{0}^{b_{p}}\mbE \left|\frac{\sum\limits_{j \in \mathcal{A}^{c}}\left\{I\left(\left|\widetilde{W}_{nj}\right|  \geqslant t\right)-\Pr\left(\widetilde{W}_{nj} \geqslant t\right)\right\}}{pG(t)}\right|^{2} dt \\
= o(v_p).
\]
Note that
$$
\begin{aligned}
&\int_{0}^{b_{p}}\mbE \left|\frac{\sum\limits_{j \in \mathcal{A}^{c}}\left\{I\left(\left|\widetilde{W}_{nj}\right|  \geqslant t\right)-\Pr\left(\widetilde{W}_{nj} \geqslant t\right)\right\}}{pG(t)}\right|^{2} dt \\
=&\int_{0}^{b_{p}}\frac{\sum\limits_{i, j \in \mathcal{B}_{0}}\left\{\Pr\left(\widetilde{W}_{ni} \geqslant t, \widetilde{W}_{nj} \geqslant t\right)-\Pr \left(\widetilde{W}_{ni} \geqslant t\right) \Pr\left(\widetilde{W}_{nj} \geqslant t\right)\right\}}{p^{2} G^{2}(t)}dt\\
&+\int_{0}^{b_{p}}\frac{\sum\limits_{i, j \in \mathcal{B}_{1}}\left\{\Pr\left(\widetilde{W}_{ni} \geqslant t, \widetilde{W}_{nj} \geqslant t\right)-\Pr \left(\widetilde{W}_{ni} \geqslant t\right) \Pr\left(\widetilde{W}_{nj} \geqslant t\right)\right\}}{p^{2} G^{2}(t)}dt\\
&+ \int_{0}^{b_{p}}\frac{\sum\limits_{j \in \mathcal{A}_{c}}\left\{\Pr\left(\widetilde{W}_{nj} \geqslant t\right)-\Pr \left(\widetilde{W}_{nj} \geqslant t\right) ^2\right\}}{p^{2} G^{2}(t)}dt\\
=:& I_0 + I_1 + I_2
\end{aligned}
$$
where
$\mathcal{B}_0  = \left\{i,j\in \mathcal{A}^c; \rho^*_{ij}\leqslant (\log p)^{-2-\epsilon_0}\right\}$ and
$\mathcal{B}_1  = \left\{i,j\in \mathcal{A}^c; \rho^*_{ij}\geqslant (\log p)^{-2-\epsilon_0}\right\}$.

We first show that $I_2 =  o(v_p)$.
$$
\begin{aligned}
I_2 
\leqslant  \int_{0}^{b_{p}}\frac{\sum\limits_{j \in \mathcal{A}^{c}}\left\{\Pr\left(\widetilde{W}_{nj} \geqslant t\right)\right\}}{p^{2} G^{2}(t)}dt
\leqslant  \int_{0}^{b_{p}} \frac{1}{pG(t)}(1+o_p(1))dt
\leqslant C\int_0^{b_p} \frac{1}{pG(t)}dt
\leqslant C\frac{b_p}{pG(b_p)}
\end{aligned}
$$

Note that $G(b_p) \geqslant C_h p^{-1}[\log p]^{h/2-d_0-1}$ by (\ref{tail2: 2}) and $b_p = O(\log p)$.
We can obtain that $\frac{b_p}{pG(b_p)}\leqslant C[\log p]^{2+d_0-h/2}$. Under the condition $2d_0<h-4$ ,
we have $I_2 =  o(v_p)$. 

As to $I_0$, it follows by Lemma 3 of \cite{cai2019large} that
\[
I_0 
      \leqslant C\int_0^{b_p} (\log p)^{-\frac{3}{2}}dt
     = O((\log p)^{-\frac{1}{2}}) = o(v_p).
\]

As to  $I_1$, it follows by Lemma 3 of \cite{cai2019large} that for any $\delta > 0$,
$$
\begin{aligned}
I_1 
& \leqslant \int_{0}^{b_{p}}\frac{\sum\limits_{i, j \in \mathcal{B}_{1}}\left\{\Pr\left(\widetilde{W}_{ni} \geqslant t, \widetilde{W}_{nj} \geqslant t\right)\right\}}{p^{2} G^{2}(t)}dt\\
& \leqslant C\int_{0}^{b_{p}}\frac{\sum\limits_{i, j \in \mathcal{B}_{1}}\left\{(t+1)^{-1}\exp(-t/(1+\rho^*_{ij}+\delta)\right\}}{p^{2} G^{2}(t)}dt
\end{aligned}
$$

Note that
$$
\begin{aligned}
I_1 &  \leqslant C \sum\limits_{i, j \in \mathcal{B}_{1}} \int_0^{b_p} \frac{\exp(-t/(1+\rho^*_{ij}+\delta))}{p^2G(t)^2}dt.
\end{aligned}
$$

Note $G(t)$ is a strictly decreasing function,
$$
\begin{aligned}
I_1 \leqslant & \ C \frac{1}{p^2G(b_p)^2}  \sum\limits_{i, j \in \mathcal{B}_{1}} \int_0^{b_p} \exp(-t/(1+\rho^*_{ij}+\delta))dt\\
\leqslant & C \frac{1}{p^2G(b_p)^2}  \sum\limits_{i, j \in \mathcal{B}_{1}} (1+\rho_{ij}^*+\delta)[1-\exp(-\frac{b_p}{1+\rho_{ij}^*+\delta})]\\
 \leqslant& C \frac{1+\rho_{ij}^*+\delta}{p^2G(b_p)^2}  |\mathcal{B}_1|= O(\frac{|\mathcal{B}_1|}{[\log p]^{h-2d_0-2}}).
\end{aligned}
$$
By Condition (D1), we have $I_1 = o(v_p)$. As a result,  (\ref{ea.11}) is valid.
We can show (\ref{ea.12}) in a similar proof to that for (\ref{ea.11}). Then, it completes the
proof of Theorem 2.

\bibliographystyle{apalike}
\bibliography{bibliography}

\begin{thebibliography}{}

\bibitem[Barber and Cand{\`e}s, 2015]{bc2015}
Barber, R.~F. and Cand{\`e}s, E.~J. (2015).
\newblock Controlling the false discovery rate via knockoffs.
\newblock {\em The Annals of Statistics}, 43(5):2055--2085.

\bibitem[Belloni et~al., 2018]{Belloni.2018}
Belloni, A., Chernozhukov, V., Chetverikov, D., and Wei, Y. (2018).
\newblock Uniformly valid post-regularization confidence regions for many
  functional parameters in z-estimation framework.
\newblock {\em Annals of Statistics}, 46(6B):3643 -- 3675.

\bibitem[Belloni et~al., 2015]{Belloni.2015}
Belloni, A., Chernozhukov, V., and Kato, K. (2015).
\newblock Uniform post-selection inference for least absolute deviation
  regression and other z-estimation problems.
\newblock {\em Biometrika}, 102(1):77--94.

\bibitem[Benjamini and Hochberg, 1995]{benjamini1995controlling}
Benjamini, Y. and Hochberg, Y. (1995).
\newblock Controlling the false discovery rate: a practical and powerful
  approach to multiple testing.
\newblock {\em Journal of the Royal statistical society: series B},
  57(1):289--300.

\bibitem[Benjamini and Yekutieli, 2001]{benjamini2001control}
Benjamini, Y. and Yekutieli, D. (2001).
\newblock The control of the false discovery rate in multiple testing under
  dependency.
\newblock {\em Annals of Statistics}, 29:1165--1188.

\bibitem[Bolstad et~al., 2003]{bolstad2003comparison}
Bolstad, B.~M., Irizarry, R.~A., {\AA}strand, M., and Speed, T.~P. (2003).
\newblock A comparison of normalization methods for high density
  oligonucleotide array data based on variance and bias.
\newblock {\em Bioinformatics}, 19(2):185--193.

\bibitem[Cai et~al., 2019]{cai2019large}
Cai, T., Cai, T.~T., Liao, K., and Liu, W. (2019).
\newblock Large-scale simultaneous testing of cross-covariance matrices with
  applications to phewas.
\newblock {\em Statistica Sinica}, 29(2):983--1005.

\bibitem[Candes et~al., 2018]{candes2018panning}
Candes, E., Fan, Y., Janson, L., and Lv, J. (2018).
\newblock Panning for gold: `model-x' knockoffs for high dimensional controlled
  variable selection.
\newblock {\em Journal of the Royal Statistical Society: Series B},
  80(3):551--577.

\bibitem[Chiang et~al., 2006]{chiang2006homozygosity}
Chiang, A.~P., Beck, J.~S., Yen, H.-J., Tayeh, M.~K., Scheetz, T.~E.,
  Swiderski, R.~E., Nishimura, D.~Y., Braun, T.~A., Kim, K.-Y.~A., Huang, J.,
  et~al. (2006).
\newblock Homozygosity mapping with snp arrays identifies trim32, an e3
  ubiquitin ligase, as a bardet--biedl syndrome gene (bbs11).
\newblock {\em Proceedings of the National Academy of Sciences},
  103(16):6287--6292.

\bibitem[Cook, 2004]{cook2004testing}
Cook, R.~D. (2004).
\newblock Testing predictor contributions in sufficient dimension reduction.
\newblock {\em The Annals of Statistics}, 32(3):1062--1092.

\bibitem[Dong et~al., 2016]{dong2016permutation}
Dong, Y., Yang, C., and Yu, Z. (2016).
\newblock On permutation tests for predictor contribution in sufficient
  dimension reduction.
\newblock {\em Journal of Multivariate Analysis}, 149:81--91.

\bibitem[Fan and Li, 2001]{fan2001variable}
Fan, J. and Li, R. (2001).
\newblock Variable selection via nonconcave penalized likelihood and its oracle
  properties.
\newblock {\em Journal of the American Statistical Association},
  96(456):1348--1360.

\bibitem[Fan et~al., 2020a]{fan2020statistical}
Fan, J., Li, R., Zhang, C.-H., and Zou, H. (2020a).
\newblock {\em Statistical Foundations of Data Science}.
\newblock Chapman and Hall/CRC.

\bibitem[Fan et~al., 2020b]{fan2019rank}
Fan, Y., Demirkaya, E., Li, G., and Lv, J. (2020b).
\newblock Rank: large-scale inference with graphical nonlinear knockoffs.
\newblock {\em Journal of the American Statistical Association},
  115(529):362--379.

\bibitem[Fang et~al., 2020]{fang2020test}
Fang, E.~X., Ning, Y., and Li, R. (2020).
\newblock Test of significance for high-dimensional longitudinal data.
\newblock {\em Annals of Statistics}, 48(5):2622--2645.

\bibitem[Fang et~al., 2017]{fang2017testing}
Fang, E.~X., Ning, Y., and Liu, H. (2017).
\newblock Testing and confidence intervals for high dimensional proportional
  hazards models.
\newblock {\em Journal of the Royal Statistical Society: Series B},
  79(5):1415--1437.

\bibitem[Hall and Li, 1993]{hall1993almost}
Hall, P. and Li, K.-C. (1993).
\newblock On almost linearity of low dimensional projections from high
  dimensional data.
\newblock {\em The Annals of Statistics}, 21:867--889.

\bibitem[Huang et~al., 2010]{huang2010variable}
Huang, J., Horowitz, J.~L., and Wei, F. (2010).
\newblock Variable selection in nonparametric additive models.
\newblock {\em Annals of Statistics}, 38(4):2282--2313.

\bibitem[Huang et~al., 2008]{huang2008adaptive}
Huang, J., Ma, S., and Zhang, C.-H. (2008).
\newblock Adaptive lasso for sparse high-dimensional regression models.
\newblock {\em Statistica Sinica}, 18:1603--1618.

\bibitem[Inglot, 2010]{inglot2010inequalities}
Inglot, T. (2010).
\newblock Inequalities for quantiles of the chi-square distribution.
\newblock {\em Probability and Mathematical Statistics}, 30(2):339--351.

\bibitem[Inglot and Ledwina, 2006]{inglot2006asymptotic}
Inglot, T. and Ledwina, T. (2006).
\newblock Asymptotic optimality of new adaptive test in regression model.
\newblock In {\em Annales de l'Institut Henri Poincare (B) Probability and
  Statistics}, volume~42, pages 579--590. Elsevier.

\bibitem[Javanmard and Montanari, 2014]{javanmard2014confidence}
Javanmard, A. and Montanari, A. (2014).
\newblock Confidence intervals and hypothesis testing for high-dimensional
  regression.
\newblock {\em The Journal of Machine Learning Research}, 15(1):2869--2909.

\bibitem[Li, 2018]{li2018sufficient}
Li, B. (2018).
\newblock {\em Sufficient Dimension Reduction: Methods and Applications with
  R}.
\newblock CRC Press.

\bibitem[Li et~al., 2017]{li2017variable}
Li, R., Liu, J., and Lou, L. (2017).
\newblock Variable selection via partial correlation.
\newblock {\em Statistica Sinica}, 27(3):983--996.

\bibitem[Liu, 2013]{liu2013gaussian}
Liu, W. (2013).
\newblock Gaussian graphical model estimation with false discovery rate
  control.
\newblock {\em The Annals of Statistics}, 41:2948--2978.

\bibitem[Loh and Wainwright, 2015]{loh2015regularized}
Loh, P.-L. and Wainwright, M.~J. (2015).
\newblock Regularized m-estimators with nonconvexity: Statistical and
  algorithmic theory for local optima.
\newblock {\em The Journal of Machine Learning Research}, 16(1):559--616.

\bibitem[Ning and Liu, 2017]{ning2017general}
Ning, Y. and Liu, H. (2017).
\newblock A general theory of hypothesis tests and confidence regions for
  sparse high dimensional models.
\newblock {\em The Annals of Statistics}, 45(1):158--195.

\bibitem[Scheetz et~al., 2006]{scheetz2006regulation}
Scheetz, T.~E., Kim, K.-Y.~A., Swiderski, R.~E., Philp, A.~R., Braun, T.~A.,
  Knudtson, K.~L., Dorrance, A.~M., DiBona, G.~F., Huang, J., Casavant, T.~L.,
  et~al. (2006).
\newblock Regulation of gene expression in the mammalian eye and its relevance
  to eye disease.
\newblock {\em Proceedings of the National Academy of Sciences},
  103(39):14429--14434.

\bibitem[Shao et~al., 2007]{shao2007marginal}
Shao, Y., Cook, R.~D., and Weisberg, S. (2007).
\newblock Marginal tests with sliced average variance estimation.
\newblock {\em Biometrika}, 94(2):285--296.

\bibitem[Shi et~al., 2019]{shi2019linear}
Shi, C., Song, R., Chen, Z., and Li, R. (2019).
\newblock Linear hypothesis testing for high dimensional generalized linear
  models.
\newblock {\em The Annals of Statistics}, 47(5):2671--2703.

\bibitem[Storey, 2002]{storey2002direct}
Storey, J.~D. (2002).
\newblock A direct approach to false discovery rates.
\newblock {\em Journal of the Royal Statistical Society: Series B},
  64(3):479--498.

\bibitem[Sun and Zhang, 2021]{sun2021targeted}
Sun, Q. and Zhang, H. (2021).
\newblock Targeted inference involving high-dimensional data using nuisance
  penalized regression.
\newblock {\em Journal of the American Statistical Association},
  116(535):1472--1486.

\bibitem[Tibshirani, 1996]{tibshirani1996regression}
Tibshirani, R. (1996).
\newblock Regression shrinkage and selection via the lasso.
\newblock {\em Journal of the Royal Statistical Society: Series B},
  58(1):267--288.

\bibitem[Van~de Geer et~al., 2014]{van2014asymptotically}
Van~de Geer, S., B{\"u}hlmann, P., Ritov, Y., and Dezeure, R. (2014).
\newblock On asymptotically optimal confidence regions and tests for
  high-dimensional models.
\newblock {\em The Annals of Statistics}, 42(3):1166--1202.

\bibitem[Vershynin, 2018]{vershynin2018high}
Vershynin, R. (2018).
\newblock {\em High-dimensional Probability: An Introduction with Applications
  in Data Science}, volume~47.
\newblock Cambridge university press.

\bibitem[Wang et~al., 2012]{wang2012quantile}
Wang, L., Wu, Y., and Li, R. (2012).
\newblock Quantile regression for analyzing heterogeneity in ultra-high
  dimension.
\newblock {\em Journal of the American Statistical Association},
  107(497):214--222.

\bibitem[Xia et~al., 2018]{xia2018joint}
Xia, Y., Cai, T.~T., and Li, H. (2018).
\newblock Joint testing and false discovery rate control in high-dimensional
  multivariate regression.
\newblock {\em Biometrika}, 105(2):249--269.

\bibitem[Yin and Cook, 2002]{yin2002dimension}
Yin, X. and Cook, R.~D. (2002).
\newblock Dimension reduction for the conditional kth moment in regression.
\newblock {\em Journal of the Royal Statistical Society: Series B},
  64(2):159--175.

\bibitem[Yu and Dong, 2016]{yu2016model}
Yu, Z. and Dong, Y. (2016).
\newblock Model-free coordinate test and variable selection via directional
  regression.
\newblock {\em Statistica Sinica}, 26:1159--1174.

\bibitem[Zhang and Zhang, 2014]{zhang2014confidence}
Zhang, C.-H. and Zhang, S.~S. (2014).
\newblock Confidence intervals for low dimensional parameters in high
  dimensional linear models.
\newblock {\em Journal of the Royal Statistical Society: Series B},
  76(1):217--242.

\end{thebibliography}

\newpage

\centerline{\large Supplement to}
\centerline{\large \bf Model-Free Statistical Inference on High-Dimensional Data}

\begin{abstract}
Section S.1 consists of some technical lemmas which are used in the proof of Theorem \ref{thm1},
Theorem \ref{thm2}, and Proposition \ref{prop1}. Section S.2 provides the proof of Proposition \ref{prop1}.
Section S.3 presents additional simulation results.
\end{abstract}

\setcounter{subsection}{0}
\setcounter{equation}{0}
\setcounter{table}{0}
\setcounter{figure}{0}

\renewcommand{\thesubsection}{S.\arabic{subsection}}
\renewcommand{\theequation}{S.\arabic{equation}}
\renewcommand{\thefigure}{S.\arabic{figure}}
\renewcommand{\thetable}{S.\arabic{table}}
\def\thesection{S}

We first present assumptions on penalty functions used in the penalized least squares in Section 3.

\noindent{\bf Assumption (P)}:
\begin{description}
    \item (P1) Assume that the penalty function $p_{\lambda}(t)$ is symmetric around zero,
    increasing on the nonnegative real line, and has a continuous derivative $p'_{\lambda}(t)$ with
    $p'_{\lambda}(0+)=\lambda L>0$.
    \item (P2) For $t>0$, the function $p_{\lambda}(t)/t$ is non-increasing in $t$.
    \item (P3) There exists a constant $\gamma>0$ such that $p_{\lambda,\gamma}(t)=p_{\lambda}(t)+\gamma t^2/2$ is convex.
\end{description}

Assumption (P) corresponds to Assumption 1 in \cite{loh2015regularized} and
is very mild. Many commonly-used penalty functions including the Lasso penalty
\citep{tibshirani1996regression} and SCAD penalty \citep{fan2001variable}
satisfy Assumption (P).

\subsection{Some useful lemmas}

\begin{lemma}\citep{loh2015regularized,ning2017general} Suppose assumption (P), conditions (B2)-(B4) are valid. $\hat{\btheta}_j$ defined in (\ref{theta}) satisfies
 $$
 \begin{aligned}
     \|\hat{\btheta}_j -\btheta_j^*\|_1 & \leqslant c \lambda_X s_X,\\
     \frac{1}{n}\sum_{i=1}^n[\bZ_{ij}^{\top}(\hat{\boldsymbol{\theta}_j}-\boldsymbol{\theta}_j^*)]^2 & \leqslant C \lambda_X^2s_X,
 \end{aligned}
 $$
 with probability at least $1-\frac{1}{p^{1+\epsilon_2}}$ where $\epsilon_2$ is a small constant. Here $c,C$ are universal constants.
\end{lemma}

\begin{lemma}
 Suppose assumption (P), conditions (B2)-(B4) are valid. With probability goes to 1, $\hat{\bgamma}_{kj}$ defined in (\ref{gamma1}) or (\ref{gamma2}) both satisfy
 $$
 \begin{aligned}
    \|\hat{\bgamma}_{kj} -\bgamma_{kj}\|_1 & \leqslant c \lambda_Y s_Y,\\
    \frac{1}{n}\sum_{i=1}^n[\bZ_{ij}^{\top}(\hat{\bgamma}_{kj}-\bgamma_{kj})]^2 & \leqslant C \lambda_Y^2s_Y,
 \end{aligned}
 $$
 uniformly for $j = 1,\cdots,p$
 \label{estrate}
\end{lemma}
\noindent{\em Proof}.
If $\hat{\bgamma}_{kj}$ is given by (\ref{gamma1}), the corresponding statements hold by lemma 1 and theorems 1 and 2 in \cite{loh2015regularized}. The uniformly property can be derived by union bound inequality.

If $\hat{\bgamma}_{kj}$ is given by (\ref{gamma2}), we observe that
$$
\|\hat{\bgamma}_{kj}-\bgamma_{kj}\|_1 \leqslant \|\hat{\bbeta}_k-\bbeta_k^0\|_1 + \|\hat{\beta}_{kj} - \beta_{kj}^0\|_1 \leqslant C' \lambda_Ys_Y,
$$
and
$$
\begin{aligned}
\frac{1}{n}\sum_{i=1}^n(\bZ_{ij}^{\top}(\hat{\boldsymbol{\gamma}}_{kj}-\boldsymbol{\gamma}_{kj}))^2 & = \frac{1}{n}\sum_{i=1}^n(\bX_{i}^{\top}(\hat{\boldsymbol{\beta}}_k-\boldsymbol{\beta}_k^0) - X_{ij}(\hat{\beta}_{kj} - \beta_{kj}^0))^2\\
& \leqslant \frac{1}{n}\sum_{i=1}^n(\bX_{i}^{\top}(\hat{\boldsymbol{\beta}}_k-\boldsymbol{\beta}_k^0))^2 + \frac{1}{n} \sum_{i=1}^n (X_{ij}(\hat{\beta}_{kj} - \beta_{kj}^0))^2\\
&\leqslant C\lambda_Y^2s_Y + (\hat{\beta}_{kj} - \beta_{kj}^0)^2\frac{1}{n} \sum_{i=1}^n X_{ij}^2\\
& = C\lambda_Y^2s_Y,
\end{aligned}
$$
where the last inequality uses the subgaussian property of $X_{ij}$ with uniform norm.
\begin{lemma}
     Let $X_{1}, \ldots, X_{n}$ be i.i.d mean 0 random variables. If there exist constants $L_{1}$ and $L_{2}$, such that $\Pr\left(\left|X_{i}\right| \geqslant x\right) \leqslant L_{1} \exp \left(-L_{2} x^{r}\right)$, for some $r>0$, then for $x \geqslant \sqrt{8 \mathbb{E}\left(X_{i}^{2}\right) / n}$
    $$
    \begin{aligned}
        \Pr\left(\left|\frac{1}{n} \sum_{i=1}^{n} X_{i}\right| \geqslant x\right) \leqslant 4 & \exp \left(-\frac{1}{8} n^{r /(2+r)} x^{2 r /(2+r)}\right) \\
        &+4 n L_{1} \exp \left(-\frac{L_{2} n^{r /(2+r)} x^{2 r /(2+r)}}{2^{r}}\right) .
    \end{aligned}
\label{genbernineq}
    $$
\end{lemma}
This lemma builds a general concentration inequality of univariate random variable. Details of proof are referred to \cite{ning2017general}.

\begin{lemma}
Suppose $X$ follows a chi-square distribution with $h$ degrees of freedom where $h\geqslant 2$. Denote $b_p = 2\log p + 2d_0\log\log p$. The tail probability $\Pr(X\geqslant b_p)$ satisfies
    \begin{align}
        \Pr(X\geqslant b_p) &\leqslant C_1\frac{\left(\log p\right)^{h/2-d_0-1}}{p} \label{tail2: 1}, \\
        \Pr(X\geqslant b_p) &\geqslant C_2\frac{\left(\log p\right)^{h/2-d_0-1}}{p} \label{tail2: 2},
    \end{align}
    with $p \to \infty$. Here $C_1$ and $C_2$ are two positive constants.
    \label{tailbound}
\end{lemma}
\noindent{\em Proof}.
    By Lemma 1 in \cite{inglot2006asymptotic}, we have
    $$
    \Pr(X \geqslant t) \leqslant \frac{1}{\sqrt{\pi}}\frac{t}{t-h+2} \mathcal{E}_h(t).
    $$
    where $\mathcal{E}_{h}(t)=\exp \left\{-\frac{1}{2}(t-h-(h-2) \log (t / h)+\log h)\right\}$. Since we consider $h$ as a fixed number and $p\to \infty$, we have
    $$
    \begin{aligned}
        \Pr(X \geqslant b_p)  & \leqslant C  \frac{\exp (h/2)}{\sqrt{h}} (\frac{b_p}{h})^{h/2-1}\exp \left(-\frac{1}{2}b_p\right)\\
        &\leqslant C C_h \frac{\left\{\log[p (\log p)^{d_0}]\right\}^{\frac{h}{2}-1}}{p(\log p)^{d_0}}.
    \end{aligned}
    $$

    Thus we have
    $$
    G(b_p)\leqslant CC_h\frac{\left(\log p\right)^{h/2-d_0-1}}{p}
    $$
    for some positive constant $C$ with $p$ goes to $\infty$.

    The proof of inequality (\ref{tail2: 2}) follows from the proposition 3.1 in \cite{inglot2010inequalities}:
    $$
    \begin{aligned}
        \Pr(X\geqslant t)\geqslant \frac{1-e^{-2}}{2} \frac{t}{t-h+2 \sqrt{h}} \mathcal{E}_{h}(t).
    \end{aligned}
    $$

    By substituting the formula of $b_p$, we have
    $$
    \begin{aligned}
        \mathcal{E}_{h}(b_p) & = \frac{\exp (h/2)}{\sqrt{h}} (\frac{b_p}{h})^{h/2-1}\exp \left(-\frac{1}{2}b_p\right)\\
        & = C_h \frac{\left\{\log[p (\log p)^{d_0}]\right\}^{\frac{h}{2}-1}}{p(\log p)^{d_0}}
    \end{aligned}
    $$
    where $C_h = \frac{\exp (h/2)}{h^{(h-1)/2}}.$ After simple calculation, we can obtain
    $$
    \begin{aligned}
        \Pr(X\geqslant b_p)\geqslant C_h \frac{(\log p)^{\frac{h}{2}-d_0-1}}{p}
    \end{aligned}
    $$
    for some constant $C_h$ with $p$ goes $\infty$.

\begin{lemma}
Suppose $X$ follows a chi-square distribution with $h$ degrees of freedom where $h\geqslant 2$. Let $G(t) = \Pr(X\geqslant t)$. For any $t \in [0,K\log p]$, the following equation
    $$
    G(t+\delta) = G(t)(1+o(1))
    $$
    holds uniformly if $\delta = o(1)$.
    \label{tailrate}
\end{lemma}
\noindent{\em Proof}.
    Note
    \begin{equation}
        \begin{aligned}
            \frac{G(t+\delta)}{G(t)} & = \frac{G(t) - \int_{t}^{t+\delta} g(x)dx}{G(t)}\\
            & = 1 - \frac{\int_t^{t+\delta} g(x)dx}{G(t)}\\
            & = 1 - \frac{g(t')\delta}{G(t)} \ \ \ \ \ \text{$t' \in [t,t+\delta]$}
        \end{aligned}
    \end{equation}
    Firstly, we consider the case $t\geqslant h-1$. From the proposition 3.1 in \cite{inglot2010inequalities}:
    $$
    \begin{aligned}
        G(t) \geqslant \frac{1-e^{-2}}{2} \frac{t}{t-h+2 \sqrt{h}} \mathcal{E}_{h}(t)
    \end{aligned}
    $$
        where $\mathcal{E}_{h}(t)=\exp \left\{-\frac{1}{2}(t-h-(h-2) \log (t / h)+\log h)\right\}$.
    Thus we have
    $$
    \begin{aligned}
    \frac{g(t')\delta}{G(t)}&\leqslant C\frac{g(t')\delta}{\mathcal{E}(t)}\\
&\leqslant C_h (t')^{h/2-1}\exp(-t'/2)\exp(t/2)t^{1-h/2}\delta\\
& = C_h\delta (\frac{t'}{t})^{h/2-1} \exp(\frac{t-t'}{2})\\
& = C_h\delta(1+o(1))
    \end{aligned}
    $$
where $C_h$ only depends on $h$.
If $t<h-1$, $G(t)\geqslant G(h-1)>0$. The result still holds by noting
$$
\begin{aligned}
    \frac{G(t+\delta)}{G(t)} \geqslant 1 - \frac{g(t')\delta}{G(h-1)}\geqslant 1 - \frac{\max\limits_{t\in (0,h-1]}g(t)}{G(h-1)}\delta = 1+o(1).
\end{aligned}
$$
The reverse direction holds by
$$
\begin{aligned}
\frac{G(t+\delta)}{G(t)} \leqslant 1 - \frac{g(t')\delta}{G(0)}\leqslant 1 - \min\limits_{t\in [0,K\log p]}g(t)\delta = 1+o(1).
\end{aligned}
$$


\subsection{\bf Proof of Proposition 2}
%
%
Due to the similarity between two procedures in (\ref{gamma1}) and (\ref{gamma2}) of
estimating $\bgamma_{kj}$, we only present the proofs that $\hat{\bgamma}_{kj}$ is given by (\ref{gamma1}).

From the proof of Theorem 1(a), it follows that
$$
\max\limits_{j\in \mathcal{A}^c}\|\bar{\boldsymbol{S}}_n^j-\widetilde{\boldsymbol{S}}_n^j\|_1
= O_p\{F(n,p,s_X,s_Y)\}
= O_p\left(\max\{s_X,s_Y\} \log p/\sqrt{n}\right).
$$

Under Condition (B3), $\max\{s_X,s_Y\}= o(\sqrt{n}/\log p)$. Thus,
$\|\bar{\boldsymbol{S}}_n^j-\widetilde{\boldsymbol{S}}_n^j\|_1=o_P(1)$. This will be  used
to prove $|W_{nj}-\widetilde{W}_{nj}| = o_p(1)$. Although this condition is enough to show $W_{nj}
\rightarrow \chi_h^2$ for the fixed $j\in \mathcal{A}^c$, it is not enough to control the false discovery
rate at pre-specified level. In the proof of this proposition,
we establish its convergence rate with respect to $(n,p,s_X,s_Y)$. Here we set $\max(s_X,s_Y) = s$.

We begin with estimation of the rate of $\|\hat{\boldsymbol{\Omega}}_j -  \boldsymbol{\Omega}_j\|_2$.
Recall that
$
\hat{\boldsymbol{\Omega}}_j = \frac{1}{n}\sum_{i=1}^n \boldsymbol{S}_{i}^j\boldsymbol{S}_{i}^{j\top}
$, we define $\widetilde{\boldsymbol{\Omega}}_j = \frac{1}{n}\sum_{i=1}^n \tilde{\boldsymbol{S}}_{i}^j
\tilde{\boldsymbol{S}}_{i}^{j\top}$, where the $k$-th component of $\tilde{\boldsymbol{S}}_i^j$
is $\tilde{S}_{ik}^j = \eta^j_{ik}(X_{ij} - \bZ_{ij}^{\top}\boldsymbol{\theta}^*_j)$, $\eta^j_{ik}
= f_k(Y_i)-\boldsymbol{Z}^\top_{ij}\boldsymbol{\gamma}_{kj}$ for $k \in  (1,\cdots,h)$.
Denote $I_{1j}= \|\hat{\boldsymbol{\Omega}}_j
-\widetilde{\boldsymbol{\Omega}}_j\|_2 $ and $I_{2j}=\|\widetilde{\boldsymbol{\Omega}}_j
- \boldsymbol{\Omega}_j\|_2$. It follows by the triangle inequality that
$
\|\hat{\boldsymbol{\Omega}}_j - \boldsymbol{\Omega}_j\|_2 \leqslant  I_{1j} + I_{2j}.
$

Since $ \|\widetilde{\boldsymbol{\Omega}}_j - \boldsymbol{\Omega}_j\|_2 \leqslant h  \|\widetilde{\boldsymbol{\Omega}}_j -
\boldsymbol{\Omega}_j\|_{\max}$ and $h$ is a fixed number by assumption, $I_{2j}$ has the same convergence rate as $ \|\widetilde{\boldsymbol{\Omega}}_j - \boldsymbol{\Omega}_j\|_{\max}$.
Denote $d_{kl}^j$ to be the $(k,l)$-element of $ \widetilde{\boldsymbol{\Omega}}_j - \boldsymbol{\Omega}_j$.
$d_{kl}^j=\frac{1}{n}\sum_{i=1}^n \{\eta^j_{ik}\eta^j_{il} (X_{ij} - \bZ_{ij}^{\top}\boldsymbol{\theta}_j^*)^2
- \omega_{kl}^j\}$, where $\omega_{kl}^j$ is the $(k,l)$-element of $\boldsymbol{\Omega}_j$.
By union bound inequality, we have $
\Pr(\|\widetilde{\boldsymbol{\Omega}}_j - \boldsymbol{\Omega}_j\|_{\max}\geqslant t)
\leqslant h^2\max_{1\leqslant k,l\leqslant h} \Pr(|d_{kl}^j|\geqslant t)$.

Note that
$$
\begin{aligned}
    &\Pr(|\eta^j_{ik}\eta^j_{il} (X_{ij} - \bZ_{ij}^{\top}\boldsymbol{\theta}_j^*)^2 - \omega_{kl}| \geqslant t)\\
      \leqslant & \Pr(|\eta^j_{ik}\eta^j_{il} (X_{ij} - \bZ_{ij}^{\top}\boldsymbol{\theta}_j^*)^2|\geqslant t - |\omega_{kl}^j|)\\
    \leqslant& \Pr(\{\eta^j_{ik}\eta^j_{il}\}^2\geqslant  t - |\omega_{kl}^j|) + \Pr( (X_{ij} - \bZ_{ij}^{\top}\boldsymbol{\theta}_j^*)^4\geqslant  t - |\omega_{kl}^j|)\\
     =& \Pr(|\eta^j_{ik}\eta^j_{il}| \geqslant \sqrt{t-|\omega_{kl}^j}|)  + \Pr( (X_{ij} - \bZ_{ij}^{\top}\boldsymbol{\theta}_j^*)^2 \ \geqslant \sqrt{t-|\omega_{kl}^j|})
\end{aligned}
$$
while $t\geqslant |\omega_{kl}^j|$. Under Condition (B2), $\eta^j_{ik}\eta^j_{il}$ and $ (X_{ij} - \bZ_{ij}^{\top}\boldsymbol{\theta}^*_j)^2$ are sub-exponential random variables with uniform bounded sub-exponential norm. Thus we can find some constants $L_{1kl}, L_{2kl}$ such that $ \Pr(|\eta^j_{ik}\eta^j_{il}| \geqslant \sqrt{t-|\omega_{kl}|}) \leqslant L_{1kl} \exp(-L_{2kl}\sqrt{t})$. While $t< |\omega_{kl}^j|$, we can select $L_{3kl}$ which satisfies $ L_{3kl} \exp(-L_{2kl}\sqrt{t}) \geqslant 1$. Similar result holds for $(X_{ij} - \bZ_{ij}^{\top}\boldsymbol{\theta}^*_j)^2$. Hence, we verified the condition of Lemma \ref{genbernineq}.

It follows by Lemma \ref{genbernineq} that
for some $t \geqslant C \sqrt{(\log np)^{5}/n}$ with large constant $C$,
$$
\Pr(\|\widetilde{\boldsymbol{\Omega}}_j - \boldsymbol{\Omega}_j\|_{\max}\geqslant t)
\leqslant 4 h^{2} \exp \left(-\frac{1}{8} n^{1 / 5} t^{2 / 5}\right)+
4 n h^{2} C_{1} \exp \left(-C_{2} n^{1 / 5} t^{2 / 5}\right) = o(\frac{1}{np})
$$
for  some constants $C_1$ and $C_2$.

Thus, $\|\widetilde{\boldsymbol{\Omega}}_j
- \boldsymbol{\Omega}_j\|_{\max}= O_p( \sqrt{(\log np)^{5}/n})$.
Since $\Pr(\|\widetilde{\boldsymbol{\Omega}}_j - \boldsymbol{\Omega}_j\|_{\max}\geqslant t)
= o(1/np)$ for fixed $j$, it is easy to show
$\max\limits_{j}\|\widetilde{\boldsymbol{\Omega}}_j - \boldsymbol{\Omega}_j\|_{\max}=
O_p(\sqrt{(\log np)^{5}/n})$ holds uniformly for $j\in \mA^c$ by the union bound inequality.

As to $I_{1j}$, it suffices to derive the rate of
$\|\hat{\boldsymbol{\Omega}}_j-\widetilde{\boldsymbol{\Omega}}_j\|_{\max}$.
By definitions,
$$
\|\hat{\boldsymbol{\Omega}}_j-\widetilde{\boldsymbol{\Omega}}_j\|_{\max} 
\leqslant 2h\|\frac{1}{n}\sum_{i=1}^n [\boldsymbol{S}_i^j - \widetilde{\boldsymbol{S}}_i^j]
\widetilde{\boldsymbol{S}}_i^{j\top}\|_{\max} + h\|\frac{1}{n}\sum_{i=1}^n [\widetilde{\boldsymbol{S}}_i^j
- \boldsymbol{S}_i^j] [\widetilde{\boldsymbol{S}}_i^j - \boldsymbol{S}^j_i]^{\top}\|_{\max} 
$$
Denote $I_{1j1}=\frac{1}{n}\sum_{i=1}^n [\boldsymbol{S}_i^j - \widetilde{\boldsymbol{S}}_i^j]
\widetilde{\boldsymbol{S}}_i^{j\top}$ and
$I_{1j2}=\frac{1}{n}\sum_{i=1}^n [\widetilde{\boldsymbol{S}}_i^j
- \boldsymbol{S}_i^j] [\widetilde{\boldsymbol{S}}_i^j - \boldsymbol{S}^j_i]^{\top}$.
Decomposing $I_{1j2}$, it follows that
$$
\|I_{1j2}\|_{\max} \leqslant C \sum_{v=1}^6 \|I_{1j2v}\|_{\max}
$$
where the $(k,l)$-element of  $I_{1j2v}$ is given by
$$
\begin{aligned}
    \{I_{1j21}\}_{kl}  &= \frac{1}{n}\sum_{i=1}^n \eta^j_{ik}\eta^j_{il} [\bZ_{ij}^{\top}(\boldsymbol{\theta}_j^*-\hat{\boldsymbol{\theta}_j})]^2\\
    \{I_{1j22}\}_{kl} & =  \frac{1}{n}\sum_{i=1}^n \eta^j_{ik}(X_{ij}-\bZ_{ij}^{\top}\boldsymbol{\theta}^*_j)(\hat{\boldsymbol{\theta}_j}-\boldsymbol{\theta}_j^*)^{\top}\bZ_{ij}\bZ_{ij}^{\top}(\hat{\boldsymbol{\gamma}}_{lj}-\boldsymbol{\gamma}_{lj})\\
    \{I_{1j23}\}_{kl} & =  \frac{1}{n}\sum_{i=1}^n \eta^j_{ik}\bZ_{ij}^{\top}(\hat{\boldsymbol{\gamma}}_{lj}-\boldsymbol{\gamma}_{lj})[\bZ_{ij}^{\top}(\hat{\boldsymbol{\theta}_j} - \boldsymbol{\theta}_j^*)]^2 \\
    \{I_{1j24}\}_{kl} & =  \frac{1}{n}\sum_{i=1}^n (X_{ij} - \bZ_{ij}^{\top}\boldsymbol{\theta}_j^*)^2(\hat{\boldsymbol{\gamma}}_{lj}-\boldsymbol{\gamma}_{lj})^{\top}\bZ_{ij}\bZ_{ij}^{\top}(\hat{\boldsymbol{\gamma}}_{kj}-\boldsymbol{\gamma}_{kj})\\
    \{I_{1j25}\}_{kl} & =  \frac{1}{n}\sum_{i=1}^n(X_{ij} - \bZ_{ij}^{\top}\boldsymbol{\theta}_j^*) [\bZ_{ij}^{\top}(\boldsymbol{\theta}_j^*-\hat{\boldsymbol{\theta}_j})](\hat{\boldsymbol{\gamma}}_{lj}-\boldsymbol{\gamma}_{lj})^{\top}\bZ_{ij}
    \bZ_{ij}^{\top}(\hat{\boldsymbol{\gamma}}_{kj}-\boldsymbol{\gamma}_{kj})\\
    \{I_{1j26}\}_{kl} & =  \frac{1}{n}\sum_{i=1}^n [\bZ_{ij}^{\top}(\hat{\boldsymbol{\theta}_j} - \boldsymbol{\theta}_j^*)]^2(\hat{\boldsymbol{\gamma}}_{lj}-\boldsymbol{\gamma}_{lj})^{\top}\bZ_{ij}
    \bZ_{ij}^{\top}(\hat{\boldsymbol{\gamma}}_{kj}-\boldsymbol{\gamma}_{kj})\\
\end{aligned}
$$

For ease of presentation, we abbreviate $1 \leqslant i\leqslant n$,$1\leqslant j\leqslant
p$,$1\leqslant k,l\leqslant h$ as $i,j,k,l$ in the rest of proof procedure, respectively.

For $I_{1j21}$, we obtain
$
    \max\limits_{1\leqslant j\leqslant p}\|I_{1j21}\|_{\max} \leqslant \max\limits_{i,j,k,l}|\eta^j_{ik}\eta^j_{jl}|
    \times \max\limits_{1\leqslant j\leqslant p}\frac{1}{n}\sum_{i=1}^n [\bZ_{ij}^{\top}
    (\boldsymbol{\theta}_j^*-\hat{\boldsymbol{\theta}_j})]^2.
$
Under Condition (B2), we have
\begin{align*}
    &\Pr(\max\limits_{j,i,k}|\eta^j_{ik} - \mu_{k}^j|>t)\leqslant nph\max\limits_{1\le k \le h} \Pr(|\eta^j_{ik}- \mu_{k}^j|>t) \leqslant 2nph\exp\left(-C\frac{t^2}{J}\right),
\end{align*}
where $\mu_k^j = \mbE(\eta_{ik}^j)$, $J=\max\limits_{j,k} = \|\eta^j_{k}- \mu_{k}^j\|^2_{\varphi_2}$.
Setting $t=C'\sqrt{\log np}$ with $C'$ being a large constant, we have
\begin{eqnarray*}
    \max_{j,i,k}|\eta^j_{ik}\eta^j_{jk}|\leqslant C\log np.
\end{eqnarray*}
with probability goes to $1$.

By Lemma \ref{estrate}, it indicates that $\frac{1}{n}\sum_{i=1}^n [\bZ_{ij}^{\top}(\boldsymbol{\theta}_j^*-\hat{\boldsymbol{\theta}_j})]^2 = O_p(\lambda_X^2s_X) = O_p(\frac{\log p}{n}s)$ uniformly holds for all $j \in (1,\cdots,p)$. Hence, we obtain
$$
\max\limits_{1\leqslant j \leqslant p}\|I_{1j21}\|_{\max}  = O_p(n^{-1}\log p\log (np) s).
$$

By the property of sub-Gaussian variables and Cauchy-Schwartz inequality, it holds that
$$
\begin{aligned}
    &\max_{1\leqslant j \leqslant p}\|I_{1j22}\|_{\max}\\
     \leqslant& \max_{i,j,k} |\eta^j_{ik}(X_{ij}-\bZ_{ij}^{\top}\boldsymbol{\theta}_j^*)|\max_{j,l}\frac{1}{n}\sum_{i=1}^n (\hat{\boldsymbol{\theta}}_j-\boldsymbol{\theta}_j^*)^{\top}\bZ_{ij}\bZ_{ij}^{\top}(\hat{\boldsymbol{\gamma}}_{lj}-\boldsymbol{\gamma}_{lj})\\
    \leqslant& \max_{i,j,k}|\eta^j_{ik}(X_{ij}-\bZ_{ij}^{\top}\boldsymbol{\theta}_j^*)|\max_{j,l}\sqrt{\frac{1}{n}
    \sum_{i=1}^n(\bZ_{ij}^{\top}(\hat{\boldsymbol{\gamma}}_{lj}-\boldsymbol{\gamma}_{lj}))^2}
    \sqrt{\frac{1}{n}\sum_{i=1}^n(\bZ_{ij}^{\top}(\hat{\boldsymbol{\theta}_j}-\boldsymbol{\theta}_j^*))^2}\\
     = &O_p(\frac{\log p\log(np)}{n}s).
\end{aligned}
$$
Similarly, we have
$$
\begin{aligned}
    \max_{1\leqslant j \leqslant p}\|I_{1j23}\|_{\max} & \leqslant \max_{i,j,k}|\eta^j_{ik}|\max_{l,j}\frac{1}{n}\sum_{i=1}^n \bZ_{ij}^{\top}(\hat{\boldsymbol{\gamma}}_{lj}-\boldsymbol{\gamma}_{lj})[\bZ_{ij}^{\top}(\hat{\boldsymbol{\theta}}_j - \boldsymbol{\theta}_j^*)]^2 \\
    & \leqslant\max_{i,j,k}|\eta^j_{ik}| \max_{j,l}|\bZ_{ij}^{\top}(\hat{\boldsymbol{\gamma}}_{lj}-\boldsymbol{\gamma}_{lj})|\frac{1}{n} \sum_{i=1}^n [\bZ_{ij}^{\top}(\hat{\boldsymbol{\theta}}_j - \boldsymbol{\theta}_j^*)]^2\\
    & \leqslant  \max_{i,j,k}|\eta^j_{ik}| \max_{j,l}\|\bZ_{ij}\|_{\infty}\max_{j,l}\|\hat{\boldsymbol{\gamma}}_{lj}-\boldsymbol{\gamma}_{lj}\|_1\frac{1}{n} \sum_{i=1}^n [\bZ_{ij}^{\top}(\hat{\boldsymbol{\theta}}_j - \boldsymbol{\theta}_j^*)]^2\\
    & = O_p\left(s^2\log (np) (\frac{\log p}{n})^{\frac{3}{2}}\right),
\end{aligned}
$$
and
$$
\begin{aligned}
    &\max_{j}\|I_{1j24}\|_{\max} \\
    \leqslant & \max_{i,j}(X_{ij}-\bZ_{ij}^{\top}\btheta_j^*)^2 \max_{j,k,l}\frac{1}{n}\sum_{i=1}^n (\hat{\boldsymbol{\gamma}}_{lj}-\boldsymbol{\gamma}_{lj})^{\top}\bZ_{ij}\bZ_{ij}^{\top}(\hat{\boldsymbol{\gamma}}_{kj}-\boldsymbol{\gamma}_{kj})\\
    \leqslant& \max_{i,j}(X_{ij}-\bZ_{ij}^{\top}\boldsymbol{\theta}_j^*)^2\max_{j,l}\sqrt{\frac{1}{n}
\sum_{i=1}^n(\bZ_{ij}^{\top}(\hat{\boldsymbol{\gamma}}_{lj}-\boldsymbol{\gamma}_{lj}))^2}
\sqrt{\frac{1}{n}\sum_{i=1}^n(\bZ_{ij}^{\top}(\hat{\boldsymbol{\theta}_j}-\boldsymbol{\theta}_j^*))^2}\\
    = &O_p(\frac{\log p\log(np)}{n}s).
\end{aligned}
$$
Moreover,
$$
\begin{aligned}
\max\limits_{1\leqslant j \leqslant p}\|I_{1j25}\|_{\max} &= O_p\left(\log (np)(\frac{s\log p}{n})^{\frac{3}{2}}\right)\\
\max\limits_{1\leqslant j \leqslant p}\|I_{1j26}\|_{\max} & = O_p\left(s^3\log (np) (\frac{\log p}{n})^2\right).
\end{aligned}
$$
In summary, we have that $\max\limits_{1\leqslant j \leqslant p}\|I_{1j2}\|_{\max} = O_p(r(n,p,s)) $, where
$$
\begin{aligned}
    r(n,p,s) = &\frac{s\log p\log (np)}{n}\vee
   s^2\log (np) (\frac{\log p}{n})^{\frac{3}{2}}
    \vee s^3\log (np) (\frac{\log p}{n})^2.
\end{aligned}
$$
Similarly, we can show that $\max\limits_{1\leqslant j \leqslant p}\|I_{1j1}\|_{\max} = O_p(\sqrt{r(n,p,s)}), $ it follows Lemma S.1.16 of \cite{fang2020test} that
$$
\|\frac{1}{n}\sum_{i=1}^n [\boldsymbol{S}_i^j - \widetilde{\boldsymbol{S}}_i^j]
\widetilde{\boldsymbol{S}}_i^{j\top}\|_2 \leqslant\left\|\frac{1}{n} \sum_{i=1}^{n}\left[\boldsymbol{S}_i^j - \widetilde{\boldsymbol{S}}_i^j\right]\left[\boldsymbol{S}_i^j - \widetilde{\boldsymbol{S}}_i^j\right]^{\top}\right\|_{2}^{1 / 2}\left\|\frac{1}{n} \sum_{i=1}^{n}  \widetilde{\boldsymbol{S}}_i^j \widetilde{\boldsymbol{S}}_i^{j\top}\right\|_{2}^{1 / 2}.
$$

Thus, we have obtained the rate of $\max\limits_{1\leqslant j \leqslant p}\|\hat{\boldsymbol{\Omega}}_j - \boldsymbol{\Omega}_j\|_{\max}$ as follows
$$
O_p\left( \sqrt{\frac{(\log np)^{5}}{n}}\vee \frac{s\log p\log (np)}{n}\vee
s^2\log (np) (\frac{\log p}{n})^{\frac{3}{2}}
\vee s^3\log (np) (\frac{\log p}{n})^2\right).
$$
Hence $\max\limits_{1\leqslant j \leqslant p}\|\hat{\boldsymbol{\Omega}}_j - \boldsymbol{\Omega}_j\|_2 \leqslant h\max\limits_{1\leqslant j \leqslant p} \|\hat{\boldsymbol{\Omega}}_j
- \boldsymbol{\Omega}_j\|_{\max} = O_p(G(n,p,s))$.

We next derive the rate of $|\widetilde{\boldsymbol{S}}_n^{j\top}\boldsymbol{\Omega}_j^{-1}\widetilde{\boldsymbol{S}}^j_n
- \bar{\boldsymbol{S}}_n^{j\top}\hat{\boldsymbol{\Omega}}_j^{-1}\bar{\boldsymbol{S}}_n^{j\top}|$ under the null hypothesis.

Firstly, we show that $\max\limits_{j\in \mathcal{A}^c}\|\widetilde{\boldsymbol{S}}_n^j\|_2^2  = O_p(\log p)$. It follows
\begin{align*}
     \Pr\left(\|\frac{\widetilde{\boldsymbol{S}}_{n}^{j}}{\sqrt n}\|_2\geqslant C\sqrt{\frac{\log p}{n}}\right)
    =o({1}/{p}).
\end{align*}
where $j \in \mathcal{A}^c$. The last equation holds due to the sub-Gaussian assumption in Condition (B2).

Furthermore,
$\|\hat{\boldsymbol{\Omega}}_j^{-1}\widetilde{\boldsymbol{S}}_n^j\|_{\infty} \leqslant
\|\hat{\boldsymbol{\Omega}}_j^{-1}\widetilde{\boldsymbol{S}}_n^j\|_2\leqslant
\|\boldsymbol{\Omega}_j^{-1}\widetilde{\boldsymbol{S}}_n^j\|_2 +
\|(\hat{\boldsymbol{\Omega}}_j^{-1}-{\boldsymbol{\Omega}_j}^{-1})\widetilde{\boldsymbol{S}}_n^j\|_2  = O_p(\log p)$
by condition (D3). Therefore,
$$
\begin{aligned}
    &\quad |\widetilde{\boldsymbol{S}}_n^{j\top}\boldsymbol{\Omega}_j^{-1}\widetilde{\boldsymbol{S}}_n^j
    - \bar{\boldsymbol{S}}_n^{j\top}\hat{\boldsymbol{\Omega}}_j^{-1}\bar{\boldsymbol{S}}_n^{j\top}|\\
    &\leqslant \|\hat{\boldsymbol{\Omega}}_j^{-1}\|_{\max}\|\widetilde{\boldsymbol{S}}_n^j - \bar{\boldsymbol{S}}_n^j\|_1^2 + 2\|\hat{\boldsymbol{\Omega}}_j^{-1}\widetilde{\boldsymbol{S}}_n^j\|_{\infty}\|\widetilde{\boldsymbol{S}}_n^j - \bar{\boldsymbol{S}}_n^j\|_1 + \|\boldsymbol{\Omega}_j^{-1}-\hat{\boldsymbol{\Omega}}_j^{-1}\|_2\|\widetilde{\boldsymbol{S}}_n^j\|_1^2\\
    & = O_p(\|\widetilde{\boldsymbol{S}}_n^j - \bar{\boldsymbol{S}}_n^j\|_1^2  \vee  \|\widetilde{\boldsymbol{S}}_n^j - \bar{\boldsymbol{S}}_n^j\|_1\log p \vee  \|\boldsymbol{\Omega}_j^{-1}-\hat{\boldsymbol{\Omega}}_j^{-1}\|_2\log p)\\
    & = O_p\left(\max(\frac{(s\log p)^2}{n}, \frac{s(\log p)^2}{\sqrt{n}},G\log p)\right),
\end{aligned}
$$
since
$$
\|\hat{\boldsymbol{\Omega}}_j^{-1} - \boldsymbol{\Omega}_j^{-1}\|_2 \leqslant \|\hat{\boldsymbol{\Omega}}_j^{-1}\|_2\|\boldsymbol{\Omega}_j^{-1}\|_2\|\hat{\boldsymbol{\Omega}}_j - \boldsymbol{\Omega}_j\|_2
= O_p( G(n,p,s)).
$$
uniformly in $j \in \mathcal{A}^c$. The last equality is due to the Weyl's inequality
$
\|\hat{\boldsymbol{\Omega}}_j^{-1}\|_2 \leqslant \{\lambda_{\min}(\boldsymbol{\Omega}_j )
-\|\hat{\boldsymbol{\Omega}}_j - \boldsymbol{\Omega}_j\|_2\}^{-1} = O_p(1).
$

\subsection{Additional simulation results}

This section presents some simulation results when the SCAD penalty function is used in the
penalized least squares procedure proposed in Section 3.

\begin{figure}
    \centering
    \includegraphics[width=1\textwidth]{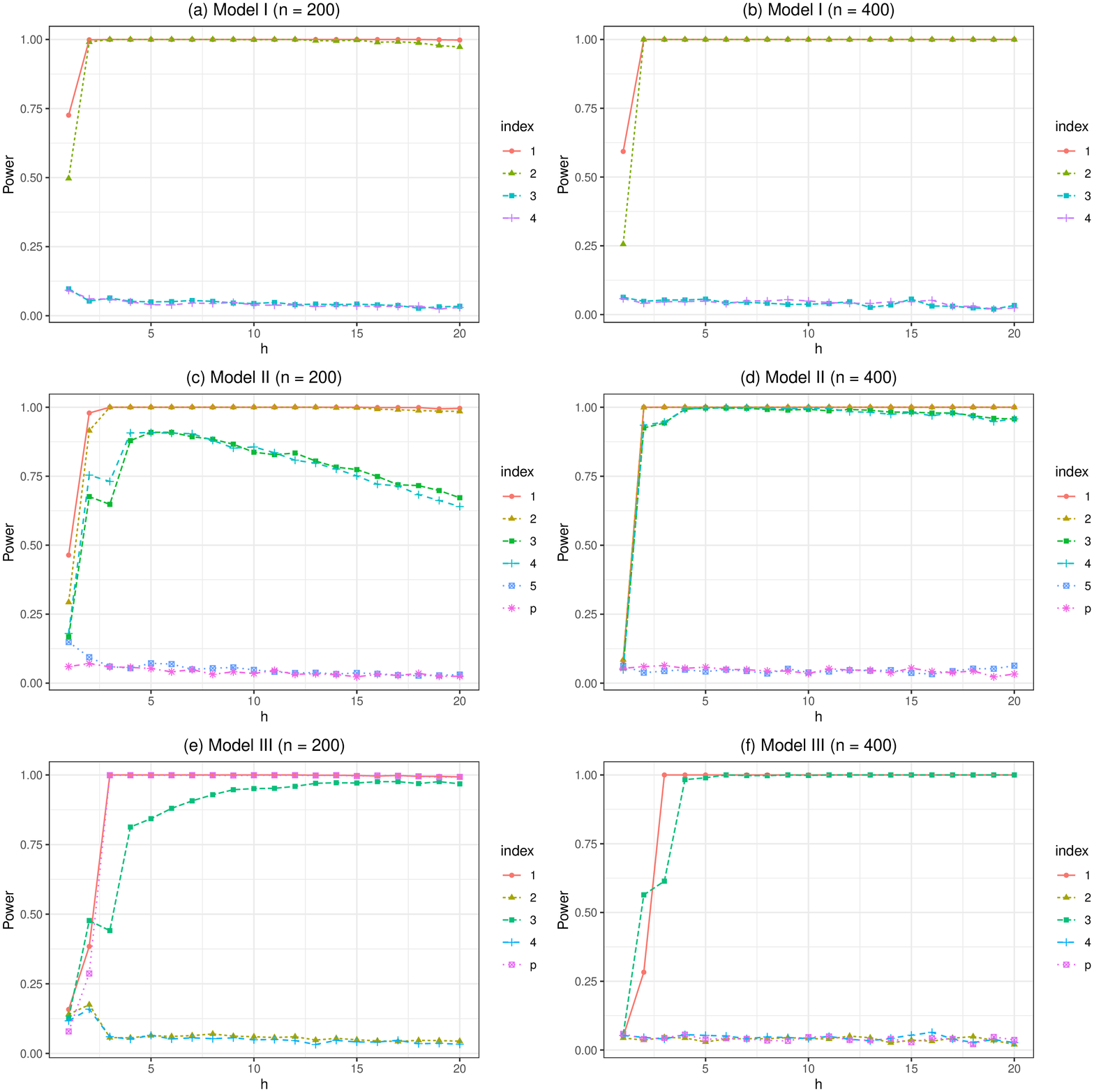}
    \caption{The plot of empirical power with respect to $h$ when the SCAD penalty is used.}
    \label{figs1}
\end{figure}

\begin{table}[htbp!]
\centering
\caption{Empirical rejection rate of $H_{0j}$ with $\alpha=5\%$ when the SCAD penalty is used. }
\label{tabs1}
\begin{tabular}{@{}cccccccccccc@{}}
\toprule
$n$ & Method         & $X_1$ & $X_2$ & $X_3$ & $X_4$ & $X_5$ & $X_{1996}$ & $X_{1997}$ & $X_{1998}$ & $X_{1999}$ & $X_{2000}$ \\
\hline
\multicolumn{12}{c}{Model I: $X_1$ and $X_2$ are active predictors }\\
\hline
200 &$W_n$-SCAD      & 1.000 & 1.000 & 0.05 & 0.040 & 0.055 & 0.037      & 0.05      & 0.049      & 0.047      & 0.05      \\
 & $T^{NL}$-SCAD   & 1.000 & 1.000 & 0.036 & 0.036 & 0.038 & 0.030      & 0.038      & 0.028      & 0.039      & 0.042      \\
\hline
400 &$W_n$-SCAD      & 1.000 & 1.000 & 0.044 & 0.056 & 0.052 & 0.050      & 0.051      & 0.045      & 0.043      & 0.056      \\
 & $T^{NL}$-SCAD   & 1.000 & 1.000 & 0.042 & 0.040 & 0.050 & 0.046      & 0.044      & 0.039      & 0.038      & 0.042      \\
\bottomrule\hline
\multicolumn{12}{c}{Model II: $X_1, X_2, X_3$ and $X_4$ are active predictors}\\
\hline
200 & $W_n$-SCAD     & 1.000                        & 1.000 & 0.909 & 0.908 & 0.072 & 0.057      & 0.042      & 0.045      & 0.042      & 0.053      \\
 & $T^{NL}$-SCAD   & 1.000                         & 1.000 & 0.062 & 0.077 & 0.031 & 0.029      & 0.038      & 0.027      & 0.043      & 0.036      \\
\hline
 & $W_n$-SCAD      & 1.000 & 1.000 & 0.998 & 0.997 & 0.047 & 0.046      & 0.060      & 0.036      & 0.054      & 0.035      \\
 & $T^{NL}$-SCAD  & 1.000                         & 1.000 & 0.130 & 0.136 & 0.039 & 0.040      & 0.050      & 0.038      & 0.049      & 0.039 \\
\bottomrule\hline
\multicolumn{12}{c}{Model III: $X_1, X_3$ and $X_p$ are active predictors}\\
\hline
200 & $W_n$-SCAD      & 1.000 & 0.064 & 0.843 & 0.065 & 0.054 & 0.043      & 0.049      & 0.045      & 0.057      & 0.999      \\
 & $T^{NL}$-SCAD   & 0.238 & 0.043 & 0.399 & 0.043 & 0.031 & 0.031      & 0.017      & 0.035      & 0.033      & 0.242      \\
\hline
400 & $W_n$-SCAD      & 1.000 & 0.031 & 0.991 & 0.041 & 0.040 & 0.040      & 0.032      & 0.049      & 0.051      & 1.000      \\
 & $T^{NL}$-SCAD   & 0.335 & 0.044 & 0.436 & 0.042 & 0.042 & 0.049      & 0.037      & 0.043      & 0.039      & 0.316      \\
\bottomrule\hline
\end{tabular}
\end{table}

\end{document}